\documentclass[twocolumn,useamsfonts,mfastrosym,usegraphicx,useepstopdf]{pasj01}
\usepackage{color}
\usepackage{comment}
\usepackage{hhline}
\usepackage{CJK}
\usepackage{morefloats}
\usepackage{graphicx}
\usepackage[toc]{appendix}
\usepackage{natbib}
\usepackage{nicefrac}

\def\mbf#1{\mbox{\boldmath ${#1}$}}
\def\Alfven{Alfv\'{e}n~}

\begin{document}

\title{Magnetohydrodynamics in a Cylindrical Shearing Box}

\author{Takeru K. Suzuki$^{1,2,3}$, Tetsuo Taki$^{4,1}$, \& Scott S. Suriano$^{1}$}
\email{stakeru@ea.c.u-tokyo.ac.jp}
\altaffiltext{1}{Graduate School of Arts \& Sciences, The University of Tokyo,
3-8-1, Komaba, Meguro, Tokyo 153-8902, Japan ; }
\altaffiltext{2}{
  Department of Astronomy, The University of Tokyo,
  7-3-1, Hongo, Bunkyo, Tokyo 113-0033, Japan 
}
\altaffiltext{3}{Komaba Institute for Science,  The University of Tokyo,
3-8-1, Komaba, Meguro, Tokyo 153-8902, Japan ; }
\altaffiltext{4}{
  Center for Computational Astrophysics, National Astronomical Observatory
  of Japan, 2-21-1, Osawa, Mitaka, 181-8588, Japan
}
\SetRunningHead{Suzuki et al.}{Cylindrical Shearing Box}


\maketitle 

\begin{abstract}
  We develop a framework for magnetohydrodynamical (MHD) simulations
  in a local cylindrical shearing box by extending the formulation of
  the Cartesian shearing box.
  We construct shearing-periodic conditions at the radial boundaries of
  a simulation box from the conservation relations of the basic MHD
  equations, taking into account the explicit radial
    dependence of physical quantities. 
  We demonstrate quasi-steady mass accretion, which cannot be handled
  by the standard Cartesian shearing box model, with an ideal MHD
  simulation in a vertically unstratified cylindrical shearing box up
  to 200 rotations. In this demonstrative run we set up
  (i) net vertical magnetic flux, (ii) a locally isothermal equation
  of state, and (iii) a sub-Keplerian equilibrium rotation, whereas
  the sound velocity and the initial \Alfven velocity have the same
  radial dependence as that of the Keplerian velocity.
  Inward mass accretion is induced to balance with the outward angular
  momentum flux of the MHD turbulence triggered by the magnetorotational
  instability in a self-consistent manner.
  We discuss detailed physical properties of the saturated magnetic field,
  in comparison to the results of a Cartesian shearing box simulation. 
\end{abstract}

\section{Introduction}
\label{sec:intro}
Accretion disks are ubiquitously formed around gravitational objects such as
black holes, neutron stars, white dwarfs, and pre-main-sequence stars.
Material in the inner part of a disk accretes onto a central object.
In order to induce the mass accretion, the angular momentum has to be
transported outward \citep{lp74}.
The molecular viscosity is insufficient to account
for the required transport rate of angular momentum, because the Reynolds
number of astrophysical objects is huge. Therefore, macroscopic processes
should operate in order to trigger mass accretion.

(Magneto)hydrodynamical ((M)HD hereafter) turbulence has been highlighted,
because it works as an effective viscosity to transport angular momentum
\citep[e.g.,][]{bh91,bh98,bn15}.
Magnetized disk winds, which remove angular momentum from a disk in the vertical
direction, have also been widely discussed \citep{bp82,pp92}.
Since these processes involve nonlinear phenomena, MHD simulations have been
performed to investigate the transfer of mass and angular momentum
\citep[e.g.,][]{haw00,mac00,pen10,li11,pb13,tom15,tak18,sur19}.

Local shearing box simulations have been widely used to examine
fine-scale MHD turbulence excited by the magnetorotational instability
\citep[MRI;][]{vel59,cha61,bh91} by zooming in on a local patch of an
accretion disk \citep{hgb95,mt95}. 
By taking into account vertical stratification \citep{bra95,sto96} the
local treatment is applied to studying the saturation of amplified magnetic
fields \citep[e.g.,][]{san04,fp07,dav10}, driving vertical outflows
and disk winds \citep{si09,bs13a,fro13,les13} and heating coronae
\citep{ms00,is14}.
It has been also extended further by including various physical
processes of radiative effects \citep{tur03,hir06,jia13}, dynamics of
dust grains \citep{joh06,tak16}, non-ideal MHD effects
\citep{ss02,san04,tun07,oh12,bs13b,sim15}, and acceleration of
high-energy particles \citep{hos15,kim16} in various types of accretion disks.
The local approach does not only apply to accretion disks but also to
proto-neutron stars that are formed through core-collapse supernovae
\citep{mas12,rem16}. 

Thus, this local approach has achieved great successes in various applications.
However, this does not mean that the local shearing box is a perfect approach.
In the shearing box approximation, the focus is small-scale phenomena,
which the curvature of a disk can be neglected, and local Cartesian
coordinates are adopted. A simulation box rotates with the equilibrium rotation
velocity at the origin of the box, and the radial direction is usually taken
as the $x$ axis. 
The Cartesian shearing coordinates have a strict symmetry across the $x=0$
plane.  While a central object is usually put in the $-x$ region, it would be
correct to regard that central object as actually located in the $+x$ region
of the same simulation, because there is no preferred direction with respect
to the $x$ axis.

Therefore, mass accretion cannot be directly captured in the Cartesian
shearing box; the integrated net mass flux across both ends of the $x$
boundaries should be strictly zero in a well constructed Cartesian shearing
box simulation.
The mass accretion rate cannot be measured directly from these simulations but
instead it is estimated from the $xy$ (radial-azimuthal) component of a stress
tensor based on the balance of angular momentum flux
(see Sub-subsection \ref{sec:angmomaccr}). 

It is key to take into account the curvature of the disk to break the $\pm x$
symmetry for a more realistic treatment. \citet{bra96} restored the terms
arising from the curvature in their Cartesian box simulation and reported that
it actually realized net mass accretion. \citet{kla03} introduced a framework of
shearing disks for their radiation HD simulation in spherical coordinates,
in which shearing periodic conditions are applied with explicit radial
dependences of physical quantities at the radial boundaries of a simulation box.
Based on this framework, \citet{obe09} performed semi-global MHD simulations
in cylindrical coordinates for the MRI in core-collapse supernovae.
While their works turned out to be great steps forward, the numerical
implementation is not still well matured; at the moment ``damping zones''
need to be prepared at the radial boundaries to suppress troublesome
oscillatory behavior of a simulation box. 

We extend the basic strategy of the shearing disk by utilizing the basis
conservation relations of mass, momentum, energy, and magnetic field in
an explicit manner. We directly apply them to shearing periodic conditions
at the radial boundaries of a local cylindrical simulation box.
Without prescribing a damping treatment at the radial boundaries,
our simulation naturally realizes the mass accretion that is balanced with the
outwardly transported angular momentum by MHD turbulence. 
We successfully incorporate the global effects, while keeping
the merits of the local approach that can capture fine-scale
turbulence in simulations that remain stable over long timescales.

We present the formulation of cylindrical shearing box simulations in
Section \ref{sec:Formulation}. The numerical implementation is described in
Section \ref{sec:setup} and Appendix \ref{sec:numrsb}. We demonstrate
one case of the simulation up to 200 rotation periods, in comparison
to results of a Cartesian shearing box, in Section
\ref{sec:results}. We discuss several future directions of our
framework in Section \ref{sec:discuss} and summarize the paper in
Section \ref{sec:sum}.  
 
\section{Cylindrical Shearing Box}
\label{sec:Formulation}
\subsection{Basic Equations}
\label{sec:beq}
We perform an MHD simulation in cylindrical coordinates, $(R,\phi,z)$,
with the rotation axis along the $z$ direction.  
The simulation box covers a region of $(R_{-}\le R\le R_{+}, \phi_{-}\le \phi
\le \phi_{+},z_{-}\le z \le z_{+})$ and rotates with the equilibrium rotation
frequency, $\mbf{\Omega}_{\rm eq,0} = \Omega_{\rm eq,0}\hat{z}$, at $R=R_{0}$
(see eqs.\ref{eq:roteq0} \& \ref{eq:Omgeq}), where 
the ``hat'' stands for a unit vector.
We usually take $R_{-} < R_{0} < R_{+}$ but $R_{0}$ does not necessarily
equal $(R_{-}+R_{+})/2$.
We restrict our simulation to regions near the midplane and neglect
the vertical component of the gravity in this paper. 
We solve MHD evolutionary equations, 
\begin{equation}
\frac{d \rho}{d t} + \rho \mbf{\nabla}\cdot \mbf{v} = 0,
\label{eq:mass}
\end{equation}
\begin{displaymath}
\rho\frac{d\mbf{v}}{dt} = -\mbf{\nabla}\left(p + \frac{B^2}{8\pi}\right) 
+ \left(\frac{\mbf{B}}{4\pi}\cdot \mbf{\nabla}\right)\mbf{B}
-\rho\frac{GM_{\star}}{R^2}\hat{R}
\end{displaymath}
\begin{equation}
 + \rho \mbf{R}\Omega_{\rm eq,0}^2 - 2\rho\mbf{\Omega}_{\rm eq,0}
 \mbf{\times}\mbf{v} 
\label{eq:mom}
\end{equation}
and
\begin{equation}
\frac{\partial \mbf{B}}{\partial t} = \mbf{\nabla \times (v\times B)}, 
\label{eq:ind}
\end{equation}
under a constraint equation of
\begin{equation}
  \mbf{\nabla\cdot B} = 0, 
  \label{eq:divB}
\end{equation}
with an isothermal equation of state,
\begin{equation}
  p=\rho c_{\rm s}^2,
  \label{eq:EoS}
\end{equation}
in the frame that rotates with $\Omega_{\rm eq,0}$. 
Here, $\rho$, $p$, $\mbf{v}$, and $\mbf{B}$ are density, gas pressure,
velocity, and magnetic field, $G$ is the gravitational constant, $M_{\star}$
is the mass of a central star, and $c_{\rm s}$ is isothermal sound speed. 
$\frac{d}{dt}$ and $\frac{\partial}{\partial t}$ denote Lagrangian and
Eulerian time derivatives, respectively. 
We adopt a locally isothermal approximation: $c_{\rm s}$ depends only on
spacial locations and does not evolve with time
(see Section \ref{sec:setup} for the detail).
We describe the numerical implementation of the gravity, the
centrifugal force, and the Coriolis force in the radial momentum
equation in Appendix \ref{sec:extforce}.

The velocity measured in this corotating frame, $\mbf{v}$ is related to
the velocity measured in the rest frame, $\mbf{u}$, via 
\begin{equation}
  \mbf{v} = \mbf{u} - R\Omega_{\rm eq,0}\hat{\phi},
\end{equation}
and therefore, the azimuthal velocity in the corotating frame is expressed as 
\begin{equation}
  v_{\phi} = R(\Omega - \Omega_{\rm eq,0}),  
\end{equation}
where $\mbf{\Omega}(R) = \Omega\hat{z}$ is the angular velocity measured
from the rest frame.

\subsection{Shearing Boundary Condition in Cylindrical Coordinates}
\label{sec:rbc}
A key in our framework of the cylindrical shearing box is how to prescribe
the shearing condition at the radial boundaries. 
We basically extend the shearing condition for Cartesian coordinates
\citep{hgb95} to cylindrical coordinates. 
In order to do so, we calculate the shear between $R_{-}$ and $R_{+}$ by
the angular difference, which gives the following shearing periodic boundary
condition for a variable, $S$: 
\begin{eqnarray}
  S(R_{\pm},\phi,z) &=& S(R_{\mp},\phi - (\Omega_{\rm eq,\pm}
  - \Omega_{\rm eq,\mp})t,z) \nonumber \\
  &=& S(R_{\mp},\phi \pm \Delta \Omega_{\rm eq}t,z),
  \label{eq:shearcd}
\end{eqnarray}
where $\Omega_{\rm eq,-}$ ($\Omega_{\rm eq,+}$) is the equilibrium angular
speed at the inner (outer) radial boundary, $R_{-}$ ($R_{+}$), and
$\Delta \Omega_{\rm eq}=\Omega_{\rm eq,-} - \Omega_{\rm eq,+}$, which is
positive for inner fast rotation. 

We need to carefully select shearing variables, $S$, from the conservation
laws of mass, momentum, energy, and magnetic field. Although the condition
of the energy is not necessary in the present paper because we assume the
locally isothermal equation of state (eq.\ref{eq:EoS}), we present the
formalism for the energy conservation for completeness. 

Conservative forms of the basic equations are presented in Appendix
\ref{sec:beq}. 
Radial differential terms in these equations should be treated with a special
care for the shearing periodic boundary condition.

\subsubsection*{Mass}
The first shearing variable is from the continuity equation (eqs.
\ref{eq:mass} and \ref{eq:massap}): 
\begin{equation}
  S_{\rm mass} = \rho v_R R, 
  \label{eq:Amass}
\end{equation}
which conserves the total mass in the simulation box. 

\subsubsection*{Radial Momentum}
The conservation of radial momentum can be realized by using radial
differential terms in eq.(\ref{eq:rmom}) of Appendix
\ref{sec:extforce}. However, we do not impose the strict conservation 
on the radial momentum flux in order to handle net mass accretion. We
start our simulation from the equilibrium profile described in
Subsection \ref{sec:init}, which indicates that the initial net radial
momentum flux is zero. 
Therefore, if we impose the conservation of the total radial momentum flux
in the simulation box, mass accretion cannot be induced, which is not
the purpose of the present work.

In order to handle mass accretion, we loosen the conservation condition.
In the shearing variable of the radial momentum flux we do not take into
account the curvature term, $u_{\phi}^2/R$, of eq.(\ref{eq:rmom}) in the
rest frame, which mostly corresponds to the centrifugal force 
in the corotating frame.
The centrifugal force is a dominant term in the radial force balance,
in addition to the gravity and the pressure gradient force. 
When the azimuthal velocity is decelerated, the inward flow of gas is
triggered. We determine the shearing condition of $v_{\phi}$ from the angular
momentum flux in order that net accretion is realized, which is
described later.  

Also, we do not consider the terms concerning $\mbf{B}$ in
eq.(\ref{eq:rmom}) in the shearing variables because the contribution
from these terms are not so significant (However, they may affect 
long-time behavior; see Appendix \ref{sec:numrsb} for the detail).

We use the radial dynamical pressure as a simple choice: 
\begin{equation}
  S_{{\rm mom},R} = \rho v_R^2 R, 
  \label{eq:Armom}
\end{equation}
In this setup, radial gas motion is not excited by the dynamical pressure
but mainly by the change of angular momentum and a small contribution
from the magnetic pressure. 



\subsubsection*{Angular Momentum}
Angular momentum flux directed to the radial direction is expressed as 
\begin{eqnarray}
  {\cal L}_{R\phi} &=& \left(\rho u_R u_{\phi} - \frac{1}{4\pi}B_R
  B_{\phi}\right) R \nonumber \\
  &=& \rho v_R R (R\Omega_{\rm eq}) + \left(\rho v_R \delta v_{\phi}  
  - \frac{1}{4\pi}B_R B_{\phi}\right)R \nonumber \\
  &\equiv& \rho v_R R (R\Omega_{\rm eq}) + w_{R\phi}R
  \label{eq:amflx}
\end{eqnarray}
in the rest frame (see eq.\ref{eq:angmom}), where
\begin{equation}
  \delta v_{\phi} = v_{\phi} - v_{\phi,{\rm eq}}
  \label{eq:dvphi}
\end{equation}
is the difference of $v_{\phi}$ from the local equilibrium azimuthal
velocity, $v_{\phi,{\rm eq}} = R(\Omega_{\rm eq} - \Omega_{\rm
  eq,0})$, and
\begin{equation}
  w_{R\phi} = \rho v_R \delta v_{\phi} - \frac{1}{4\pi}B_R B_{\phi}
  \label{eq:wrp}
\end{equation}
is the $R\phi$ component of MHD stress tensor.
$w_{R\phi}$ is often discussed in terms of the $\alpha$ prescription
\citep{ss73} as $w_{R\phi} = \alpha \rho c_{\rm s}^2$. 

The first term on the right-hand side denotes the angular momentum
advected by radial mass flow, and the second term corresponds to
the angular momentum transported by MHD turbulence.
When the mass accretes inward by the outward transport of angular momentum
by turbulence as in standard accretion disks \citep{ss73}, the first term
is negative and the second term is positive. 

The difference between ${\cal L}_{R\phi}R$ at $R_{-}$ and
${\cal L}_{R\phi}R$ at $R_{+}$ determines the variation of the total
angular momentum in the simulation box (eq.\ref{eq:angmom}). If
$({\cal L}_{R\phi}R)_{+} = ({\cal L}_{R\phi}R)_{-}$ is imposed, 
the total angular momentum is conserved. In this case the radial force
balance is maintained because the centrifugal force, which balances
with the gravity and the pressure gradient force, does not change with
time. Therefore, if $({\cal L}_{R\phi}R)_+ = ({\cal L}_{R\phi}R)_-$ is
applied, mass does not accrete, which is not what we want to model. 

We allow the change of the total angular momentum in order to generate
net mass accretion. However, after the magnetic field is amplified
to be in the quasi-saturated state, time-steady mass accretion should be
realized. In order to fulfill these conflicting demands, we will have to
prescribe a shearing boundary condition that satisfies
$({\cal L}_{R\phi})_+ \approx ({\cal L}_{R\phi})_-$ after the saturated
state is achieved, while we have to loosen the strict conservation
constraint, $({\cal L}_{R\phi}R)_{+} = ({\cal L}_{R\phi}R)_{-}$. 

The first term on the right-hand side, $\rho v_R R(R\Omega_{\rm eq})$,
has a negative value and is proportional to $(R\Omega_{\rm eq})$ in
the steady accretion phase, $\rho v_{R}R(<0)=$ const. (eq.\ref{eq:Amass}).
On the other hand, the second term, $w_{R\phi}R$, is positive, and
if $w_{R\phi}R = -\rho v_R R(R\Omega_{\rm eq})$ at $R=R_{\pm}$, the gas in
the simulation box does neither gains nor loses angular momentum.

Based on this consideration, we adopt a shearing variable for the
angular momentum,
\begin{equation}
  S_{\rm mom,\phi} = w_{R\phi}/\Omega_{\rm eq}. 
  \label{eq:Aangmom}
\end{equation}
Although this choice allows the gain or loss of the angular momentum,  
in the steady-state accreting phase of $\rho v_R R=$const., it gives
$w_{R\phi}R^2$ \& ${\cal L}_{R\phi}R \propto R^2\Omega_{\rm eq}$
(eq.\ref{eq:amflx}).  
We note that $|{\cal L}_{R\phi}R|$ is an increasing function of $R$ because
specific angular momentum, $R^2\Omega_{\rm eq}$, increases with $R$.
(If this is not satisfied, the system is dynamically unstable, since
it breaks the Rayleigh's stability criterion.)

Let us consider a case in which ${\cal L}_{R\phi}R$ is negative. In this
case the total angular momentum increases because the angular
momentum that flows out of $R_-$ is smaller than the incoming angular
momentum from $R_+$. As a result, the mass accretion is eventually
reduced ($v_R$ increases), which increases ${\cal L}_{R\phi}R$.
On the other hand, if ${\cal L}_{R\phi}R$ is positive, the total
angular momentum decreases because the angular momentum that flows out
of $R_+$ is larger than the incoming angular momentum from
$R_-$. Hence, mass accretion is eventually increased ($v_{R}$
decreases), and ${\cal L}_{R\phi}R$ declines.  

We expect that the choice of eq.(\ref{eq:Aangmom}) leads to
${\cal L}_{R\phi}R\approx 0$ in a self-regulating manner after the
different components of ${\cal L}_{R\phi}R$ are canceled out. However,
this argument is based on our theoretical consideration, and hence, we
have to check whether this self-regulation is actually realized by
numerical simulation.

We describe our specific method for how to numerically prescribe the
shearing condition of $S_{\rm mom,\phi}$ in Appendix \ref{sec:shstg}. 
In short, we assume both the Reynolds and Maxwell stresses have the same
scaling on $R$, $\rho v_R \delta v_{\phi}\propto \Omega_{\rm eq}$ and
$B_RB_{\phi}\propto \Omega_{\rm eq}$. 
The condition for the Reynolds stress gives
\begin{equation}
  \delta v_{\phi} \propto R\Omega_{\rm eq}
  \label{eq:dvphOmg}
\end{equation}
for the constraint of mass conservation, $\rho v_R\propto R^{-1}$
(eq.\ref{eq:Amass}). 

\subsubsection*{Vertical Momentum}
The shearing condition for vertical velocity is obtained from the
radial differential terms of eq.(\ref{eq:zmom}). Here, we use the only
HD term and neglect the magnetic effect
($\partial_{R}(B_RB_zR)$), because the latter is generally small in
the unstratified setting.  We use
\begin{equation}
  S_{{\rm mom},z}=\rho v_{R}v_{z}R.
  \label{eq:Azmom}
\end{equation}
as a shearing variable for the vertical momentum. 

\subsubsection*{Magnetic Field}
Similarly to the HD variables, the radial differential
terms of the induction equation should be used as shearing variables,
which are the $z$ component of induction electric field, 
\begin{equation}
  S_{B_{\phi}} = cE_{z} = v_R B_{\phi} - v_{\phi} B_R,
  \label{eq:ABp}
\end{equation}
in the evolutionary equation of $B_{\phi}$ (eq.\ref{eq:indphi}), and
the $\phi$ component of induction electric field,  
\begin{equation}
  S_{B_z} = R c E_{\phi} = R (v_z B_R - v_R B_z),
  \label{eq:ABz}
\end{equation}
in the evolutionary equation of $B_{z}$ (eq.\ref{eq:indz}), where $c$ is
the speed of light. 
Here, we note that the induction equations in the corotating frame can be
derived by replacing $u_{\phi}$ with $v_{\phi}$ of
eqs.(\ref{eq:indR}) -- (\ref{eq:indz}) in the rest frame (see Appendix
\ref{sec:corrst}). 
Besides these two shearing variables, the constraint equation
(\ref{eq:divB}) determines the three components of magnetic field. 

\subsubsection*{Energy}
The radial differential term of the total energy equation
(\ref{eq:toteng}) can be used as a shearing variable for energy.
Because the contribution from magnetic field is usually small, 
\begin{equation}
  S_{\rm eng} = \rho v_R R \left[\frac{v^2}{2} + (\gamma - 1)e \right]
  \label{eq:Aeng}
\end{equation}
can be a reasonable shearing variable, where $\gamma$ is the ratio of
specific heats and we used the relation, $p=(\gamma - 1)\rho e$. 

However, as we stated above, we assume that the gas is locally isothermal (see
Section \ref{sec:setup} for the detail) and we do not solve the energy equation.
Therefore, we do not use $S_{\rm eng}$ for the radial shearing boundary
condition in this paper.

\subsubsection*{Summary of Shearing Variables}
We set up the seven shearing variables, eqs.(\ref{eq:Amass}),
(\ref{eq:Armom}), (\ref{eq:Aangmom})--(\ref{eq:Aeng}). 
The eight primitive variables, $\rho$, $\mbf{v}$, $\mbf{B}$, and
$e$ for the shearing periodic condition are in principle determined by these
seven conditions and the constraint of $\mbf{\nabla\cdot B}=0$
(eq.\ref{eq:divB}). A specific implementation method needs to be carefully
constructed in order that it is compatible with an adopted MHD scheme. 
We describe our method in Subsection \ref{sec:ourbc} and Appendix
\ref{sec:shstg}.

\subsection{Periodic Boudary for $\phi$ and $z$ Components}
\label{sec:pzbc}
We adopt the periodic boundary condition for a variable, $A$,
at the $\phi$ and $z$ boundaries, as usually taken in unstratifed Cartesian
shearing box simulations \citep[][and more]{hgb95}, 
\begin{equation}
A(R,\phi_{\pm},z) = A(R,\phi_{\mp},z)
\end{equation}
and
\begin{equation}
A(R,\phi,z_{\pm}) = A(R,\phi,z_{\mp}).
\end{equation}
For the $\phi$ and $z$ boundaries, we take primitive variables for
$A = \rho, \mbf{v}, \mbf{B}$, and $e$.

\subsection{Constraints \& Conserved Quantities}
\label{sec:cq}
We can obtain constraints and conserved quantities from the shearing
periodic boundary condition for the $R$ direction (Subsection
\ref{sec:rbc}) and the simple periodic boundary condition for the
$\phi$ and $z$ directions 
(\S \ref{sec:pzbc}). 
The shearing condition of $S_{\rm mass}$ (eq.\ref{eq:Amass}) ensures
the conservation of the mass in the simulation box
\begin{equation}
  M=[\rho]_{_V} = \int_{z_{-}}^{z_{+}} \int_{\phi_{-}}^{\phi_{+}}\int_{R_{-}}^{R_{+}}
  \rho R dR d\phi dz = {\rm const.}, 
  \label{eq:MassConsv}
\end{equation}
where $[ \cdots ]_{V} \equiv \int_{V} dV$ represents
the volumetric integral in the entire box.  
The vertical momentum flux integrated in the box has an upper
  bound, 
\begin{eqnarray}
  [\rho v_z]_{_V} &=& \int_{z_{-}}^{z_{+}} \int_{\phi_{-}}^{\phi_{+}}
  \int_{R_{-}}^{R_{+}} (\rho v_z) R dR d\phi dz \nonumber \\
  &<& \left| \int_{z_{-}}^{z_{+}}\int_{\phi_{-}}^{\phi_{+}}d\phi dz
    \left[\frac{B_R B_{z}R}{4\pi}\right]_{R_{-}}^{R_{+}}\right|, 
\end{eqnarray}
from eqs.(\ref{eq:Azmom}) and (\ref{eq:zmom}).
The contribution from the Lorentz force (the right-hand side) is
generally small, and therefore, $[\rho v_z]_{_V}\approx 0$ is also an
approximately conserved quantity.

As we explained in Subsection \ref{sec:rbc}, we do not conserve the integrated
radial or angular momentum in order to handle net mass accretion.
Instead, we can derive the equations that describe epicyclic oscillations,
similarly to those obtained in the Cartesian coordinates \citep[e.g.,][]{hgb95}.
If we neglect the magnetic terms, 
by integrating the $R$ and $\phi$ components of eq.(\ref{eq:mom}) we
approximately have  
\begin{equation}
  \frac{\partial}{\partial t}[ \rho v_{R} ]_{_V} \approx 2\Omega_{\rm eq,0}
       [\rho \delta v_{\phi}]_{_V}
  \label{eq:epiR}
\end{equation}
and
\begin{equation}
  \frac{\partial}{\partial t}[\rho \delta v_{\phi} R ]_{_V}
  \approx -\frac{1}{2}\Omega_{\rm eq,0} [\rho v_{R} R]_{_V},
  \label{eq:epiphi}, 
\end{equation} 
where we assumed that $\Omega_{\rm eq}(R)$ is roughly proportional to
$R^{-3/2}$, which is valid for the thin disk condition (see Section
\ref{sec:setup}). The detailed derivations of eqs.(\ref{eq:epiR}) \&
(\ref{eq:epiphi}) are described in Appendix \ref{sec:epiosc}.

The periodic $\phi$ and $z$ boundaries guarantee the conservation of the
radial magnetic flux,  
\begin{equation}
  \Phi_R = \int_{z_{-}}^{z_{+}} \int_{\phi_{-}}^{\phi_{+}} B_R R d\phi dz,
\end{equation}
at any $R$ plane, which is independent from the radial shearing
boundary. 

The azimuthal magnetic flux, 
\begin{equation}
  \Phi_{\phi} = \int_{z_{-}}^{z_{+}} \int_{R_{-}}^{R_{+}} B_{\phi} dR dz,
  \label{eq:phiphi}
\end{equation}
is conserved from eq.(\ref{eq:ABp}) at {\it shearing planes}, which
are defined at $\phi = (\Omega_{\rm eq}(R) - \Omega_{\rm eq,0})t$. 

The shearing condition of $S_{B_z}$ (eq.\ref{eq:ABz}) conserves the
vertical magnetic flux, 
\begin{equation}
  \Phi_z = \int_{\phi_{-}}^{\phi_{+}} \int_{R_{-}}^{R_{+}} B_z R dR d\phi,
  \label{eq:phiz}
\end{equation}
at any $z$ plane.

\section{Simulation Setup}
\label{sec:setup}
\begin{table*}
  \begin{center}
    \begin{tabular}{|c|c|ccc|ccc||c|}
      \hline
      \multicolumn{9}{|c|}{Cylindrical Shearing Box}\\
      \hline
      $\beta_{z,{\rm init}}$ & $H_0/R_0$ &
      \multicolumn{3}{|c|}{Simulation Region [Box Size]} &
      \multicolumn{3}{c||}{Resolution} & $\alpha_{\rm M}$
      \\
      & & $R$ & ${\phi}$ & $z$ & $N_R$ & $N_{\phi}$ & $N_z$ &
        {\scriptsize $(0.92<R/R_0<1.12)$}\\
        \hline
        $10^3$ & $0.1$ & $0.82R_0-1.22R_0$ [$4H_0$] &
        $0-\pi/6$ [$(5\pi/3)H_0$] & $\pm 0.05R_0$ [$H_0$] & $256$ & $256$
        & $64$ & 0.106 \\
        \hline
        \hline
    \end{tabular}
    \caption{Simulation parameters of the cylindrical case. The last
      column presents the Maxwell stress (eq.\ref{eq:aM}) averaged 
      over 50 -- 200 rotations in the region of $0.92R < R_0 <
      1.12R$. 
      \label{tab:cyl}}
    \end{center}
\end{table*}

\begin{table*}
  \begin{center}
    \begin{tabular}{|c|ccc|ccc||c|}
      \hline
      \multicolumn{8}{|c|}{Cartesian Shearing Box}\\
      \hline
      $\beta_{z,{\rm init}}$ & \multicolumn{3}{|c|}{Box Size} &
      \multicolumn{3}{c||}{Resolution} & $\alpha_{\rm M}$\\
      & $x$ & $y$ & $z$ & $N_x$ & $N_y$ & $N_z$ & {\scriptsize $(-H_0<x<H_0)$ }\\
      \hline
      $10^3$ & $4H_0$ & $(5\pi/3)H_0$ & $H_0$ & $256$ & $256$ & $64$ & 0.109\\
      \hline
    \end{tabular}
    \caption{Simulation parameters of the Cartesian case.
      The last column presents the Maxwell stress (eq.\ref{eq:aM}) averaged
      over 50 -- 200 rotations in the region of $-H_0 < x < H_0$. 
      \label{tab:car}}
  \end{center}
\end{table*}

The MHD simulation is performed in a vertically unstratified 
radially periodic shearing cylinder with net vertical magnetic 
fields, by neglecting the vertical component of the gravity of a
central star (eq.\ref{eq:mom}).  

\subsection{Temperature Profile}

We do not solve the energy equation (eq.\ref{eq:toteng}) but assume
an isothermal equation of state (eq.\ref{eq:EoS}). 
On the other hand, we explicitly consider the radial dependence of
temperature ($\propto c_{\rm s}^2$) in a power-law manner with a
constant index, $q_{\rm T}$, 
\begin{equation}
  c_{\rm s}^2=c_{\rm s,0}^2\left(\frac{R}{R_0}\right)^{-q_{T}}. 
  \label{eq:csraddep}
\end{equation}
This temperature profile is preserved during the simulation for the locally
isothermal assumption.  

We adopt $q_T = 1$ for the demonstrative simulation in this paper. 
This choice gives $c_{\rm s}\propto R^{-1/2}$, which is the same
scaling as that of the Keplerian rotation velocity, $R\Omega_{\rm
  K}$. We note that $q_T$, which is determined by thermal processes in
a disk, generally takes various different values under different
physical conditions. For example, $q_T=1/2$ is derived when an
accretion disk is optically thin and the temperature is determined by
the irradiation from a central star \citep[e.g.][]{hay81}; $q_{R}=3/4$
is given for a standard accretion disk, in which viscous heating is
balanced with blackbody radiation from the surfaces
\citep[e.g.,][]{pri81}. In forthcoming papers, we perform simulations
with these different $q_T$'s. 

\subsection{Simulation Region \& Resolution}
We consider a thin disk condition with the sound speed,
$c_{\rm s,0}=0.1R_0 \Omega_{\rm K}$, at $R=R_0$, where $\Omega_{\rm
  K}=\sqrt{\frac{GM_{\star}}{R^3}}$ is the Keplerian frequency.  
The scale height at $R=R_0$ can be defined as $H_0 = c_{\rm s,0}
/\Omega_{\rm K}$, which gives $H_0/R_0 = 0.1$.  
To be consistent with this approximation, we focus on a region near the
midplane and adopt a small vertical box size, $L_z = 0.1R_0 = H_0$.

We set up a larger radial box size, $L_R = 0.4R_0 = 4H_0$. The radial
spacing, $\Delta R$, of grid cells is prepared in proportion to $R$.
We use the same number of radial grid points ($=128$) inside and outside
$R=R_0$.
These settings give a radial box covering $R_{-}=0.82R_0$ to $R_{+}=1.22R_0$. 

We adopt $\pi/6$ for the azimuthal extent of the simulation box. 
The azimuthal length at $R=R_0$ of this 
case is $L_{\phi}=(5\pi/3)H_0 \approx 5.2H_0$. We also perform a
simulation in a Cartesian shearing box with the same box size to
this cylindrical 
case to inspect the effect of the different geometries.  

We resolve $H_0$ by 64 grid points in the $R$ and $z$ components. A slightly
lower resolution ($49/H_0$)
is used for the $\phi$ component. We summarize these parameters of
the cylindrical and Cartesian shearing box simulations in Tables
\ref{tab:cyl} and \ref{tab:car}, respectively.

\subsection{Radial Scalings for Shearing Periodic Boundary}
\label{sec:ourbc}

We use the six shearing variables, eqs.(\ref{eq:Amass}),
(\ref{eq:Armom}), \& (\ref{eq:Azmom}) -- (\ref{eq:ABz}), for the
radial shearing periodic condition in principle.  
However, we find $\rho$, $v_R$, and $v_z$ have simple radial dependencies
from eqs.(\ref{eq:Amass}), (\ref{eq:Armom}), \& (\ref{eq:Azmom}) under
the unstratified setup:
\begin{equation}
  \left(\frac{\rho_{+}}{\rho_{-}}\right) =
  \left(\frac{R_{+}}{R_{-}}\right)^{-q_{\rho}}
  \label{eq:smplsc1}
\end{equation}
{\bf with $q_{\rho} = 1$ and}
\begin{equation}
  \left(\frac{v_{R,+}}{v_{R,-}}\right) =
  \left(\frac{v_{z,+}}{v_{z,-}}\right) =
  \left(\frac{R_{+}}{R_{-}}\right)^0 = {\rm const}.   
  \label{eq:smplsc2}
\end{equation}


The other variables, $v_{\phi}$ and the three components of $\mbf{B}$, 
are determined from the three shearing variables, eqs (\ref{eq:Aangmom}),
(\ref{eq:ABp}) \& (\ref{eq:ABz}), and $\mbf{\nabla\cdot B}=0$
(eq.\ref{eq:divB}).
In our simulation we use staggered meshes for the HD and
magnetic field variables for the constrained transport method \citep{eh88}
to ensure $\mbf{\nabla\cdot B}=0$ (eq.\ref{eq:divB}). We describe how to
apply the shearing periodic condition on the staggered meshes in Appendix
\ref{sec:shstg}.

\subsection{Initial Condition}
\label{sec:init}
We set up a power-law dependence of the initial density
on $R$ to be consistent with the radial boundary condition of
  eq.(\ref{eq:smplsc1}) with $q_{\rho}=1$: 
\begin{equation}
  \rho_{\rm init}=\rho_{0,{\rm
      init}}\left(\frac{R}{R_0}\right)^{-q_{\rho}}. 
  \label{eq:rhoraddep}
\end{equation}
We also set a weak vertical magnetic field of 
\begin{equation}
  B_{z,{\rm init}}=B_{z,0,{\rm init}}\left(\frac{R}{R_0}\right)^{-q_{B}}, 
  \label{eq:Braddep}
\end{equation}
and the other components of magnetic field are zero, $B_R=B_{\phi}=0$.

\begin{table}[h]
  \begin{center}
  \begin{tabular}{ccc}
    \hline
    $q_{\rm T}$ & $q_{\rho}$ & $q_{B}$ \\
    \hline
    1 & 1 & 1\\
    \hline
  \end{tabular}
  \caption{Adopted power-law indices of the temperature, the density,
    and the initial vertical magnetic field, respectively.
    \label{tab:pls}}
  \end{center}
\end{table}

The initial plasma $\beta$ value is set to a constant,  
\begin{equation}
\beta_{z,{\rm init}} = 8\pi \rho c_{\rm s}^2 / B_{z,{\rm init}}^2 = 10^3, 
\label{eq:btzint}
\end{equation}
in the entire simulation box; this can be realized when the adopted power-law
indices
(eqs.\ref{eq:csraddep}, \ref{eq:rhoraddep}, \& \ref{eq:Braddep}) satisfy
$2q_B = q_{\rho} + q_T$. The present setup of $q_T=q_{\rho}=1$ gives
$q_B=1$ (Table \ref{tab:pls}).
We also note that these power-law indices give the same radial scaling
of the initial \Alfven velocity ($v_{{\rm A},z,{\rm init}}
=B_{z,{\rm init}}/\sqrt{4\pi \rho}$) $\propto R^{-1/2}$ as that of 
$c_{\rm s}$ and $R\Omega_{\rm K}$.

The equilibrium profile of the angular frequency, $\Omega_{\rm eq}$,
is derived from the radial force balance,
\begin{equation}
  R\Omega_{\rm eq}^2 - \frac{GM_{\star}}{R^2}
  - \frac{1}{\rho}\frac{\partial p}{\partial R}= 0, 
  \label{eq:roteq0}
\end{equation}
where we neglected the effect of magnetic pressure by $B_z$ because we
put very weak initial fields in our simulation. 
Because of the pressure-gradient force, $\Omega_{\rm eq}$ deviates from
$\Omega_{\rm K}$. For a positive $q_{\rho}+q_T$, the equilibrium
rotation is sub-Keplerian, $\Omega_{\rm eq} < \Omega_{\rm K}$, and we
define a sub-Keplerian parameter \citep[e.g.,][]{nak86}, 
\begin{equation}
  \eta = -\frac{1}{\rho}\frac{dp}{dR} \bigg/ 2R\Omega_{\rm K}^2 = \frac{(q_{\rho}
    +q_T)c_{\rm s}^2}{2R^2\Omega_{\rm K}^2}. 
  \label{eq:eta}
\end{equation}
Substituting eq.(\ref{eq:eta}) into eq.(\ref{eq:roteq0}), we obtain
\begin{equation}
  \Omega_{\rm eq} = \Omega_{\rm K}\sqrt{1-2\eta}. 
  \label{eq:Omgeq}
\end{equation}
The adopted $q_{T}$ and $q_{\rho}$ with $H_0/R_0=0.1$ gives the sub-Keplerian
parameter (eq.\ref{eq:eta}), $\eta \approx 0.01$. 

The wavelength, $\lambda_{\rm max,init}$, of the most unstable mode of the MRI
is derived from eqs.(\ref{eq:rhoraddep}) and (\ref{eq:Braddep}) as
\begin{eqnarray}
  \lambda_{\rm max,init} &\approx& 2\pi\sqrt{\frac{16}{15}}
  \frac{v_{{\rm A},z,{\rm init}}}{\Omega_{\rm K}} \nonumber \\
  &=& 0.029R\left(\frac{\beta_{z,{\rm init}}}{10^3}\right)^{-1/2} 
  \left(\frac{c_{\rm s,0}}{0.1R_0\Omega_{\rm K}}\right) \nonumber\\
  &=& 0.29H_0 \left(\frac{R}{R_0}\right)
  \left(\frac{\beta_{z,{\rm init}}}{10^3}\right)^{-1/2}
  \left(\frac{c_{\rm s,0}}{0.1R_0\Omega_{\rm K}}\right),
  \label{eq:lambdamax}
\end{eqnarray}
where we used the expression of the Keplerian rotation \citep[][]{bh98},
because the equilibrium rotation profile is nearly the Keplerian one with
the small sub-Keplerian index, $\eta\approx 0.01$. 

Eq. (\ref{eq:lambdamax}) shows that $\lambda_{\rm max,init}$ is proportional to
$R$; $\lambda_{\rm max,init}(R_{-}) \approx 0.24H_0$ at the inner radial
boundary of the simulation box and $\lambda_{\rm max,init}(R_{+}) \approx
0.35H_0$ at the outer boundary. $L_z$ covers 3-4 times $\lambda_{\rm max,init}$,
and therefore $\lambda_{\rm max,init}$ can be resolved by 16-22 grid points.

We add random velocity perturbations with $10^{-4}c_{\rm s}$ to the $R$ and
$\phi$ components of the equilibrium velocity distribution, $v_R=v_z=0$
and $v_{\phi} = R(\Omega_{\rm eq} - \Omega_{\rm eq,0})$, which eventually
trigger the MRI.

\subsection{Scheme}
We adopt the 2nd order Godunov + CMoCCT method to update the physical
variables \citep{san99}. In this scheme, we split the time-updating
procedure into compressible and incompressible parts; in the former we
solve the hydrodynamics with magnetic pressure by the nonlinear Godunov
method; in the latter we solve magnetic tension force by the consistent
method of characteristics \citep[][CMoC]{cl96} with the constrained
transport (CT) to ensure $\mbf{\nabla\cdot B}=0$ \citep{eh88}.

\subsection{Simulation Units}
We adopt the simulation units normalized by $R_0=1$, $\rho_0 =1$, and
$\Omega_{\rm K,0}=1$, where $\Omega_{\rm K,0}$ is the Keplerian rotation
frequency at $R=R_0$.
The velocity is normalized by $R_0\Omega_{\rm K,0}$.
The magnetic field is normalized by $R_0\Omega_{\rm K,0}\sqrt{4\pi \rho_0}$,
which deletes the $\sqrt{4\pi}$ factor in the cgs-Gauss units.

In this paper we conventionally call $2\pi/\Omega_{\rm K,0}$ ``one
rotation'' from now on, while in a strict sense,
one rotation at $R=R_0$ is $2\pi/\Omega_{\rm eq,0}(\approx
1.01\times 2\pi/\Omega_{\rm K,0})$ in the sub-Keplerian background.

\section{Results}
\label{sec:results}
We run both cylindrical and Cartesian simulations in Tables \ref{tab:cyl}
and \ref {tab:car} until 200 rotations, $t=200(2\pi/\Omega_{\rm K,0})$.


\subsection{Time Evolution} 
\begin{figure*}
  \begin{center}
    \vspace{-0.6cm}
    \includegraphics[width=0.45\textwidth]{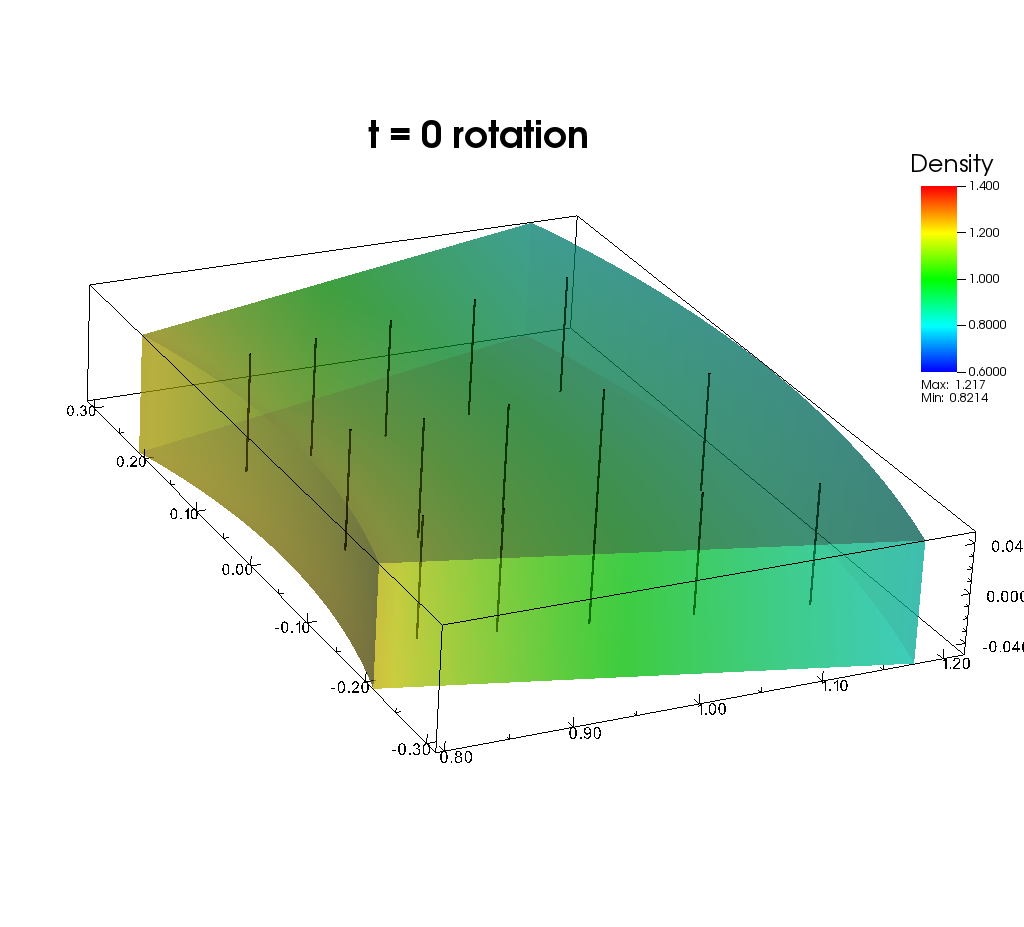}
    \includegraphics[width=0.45\textwidth]{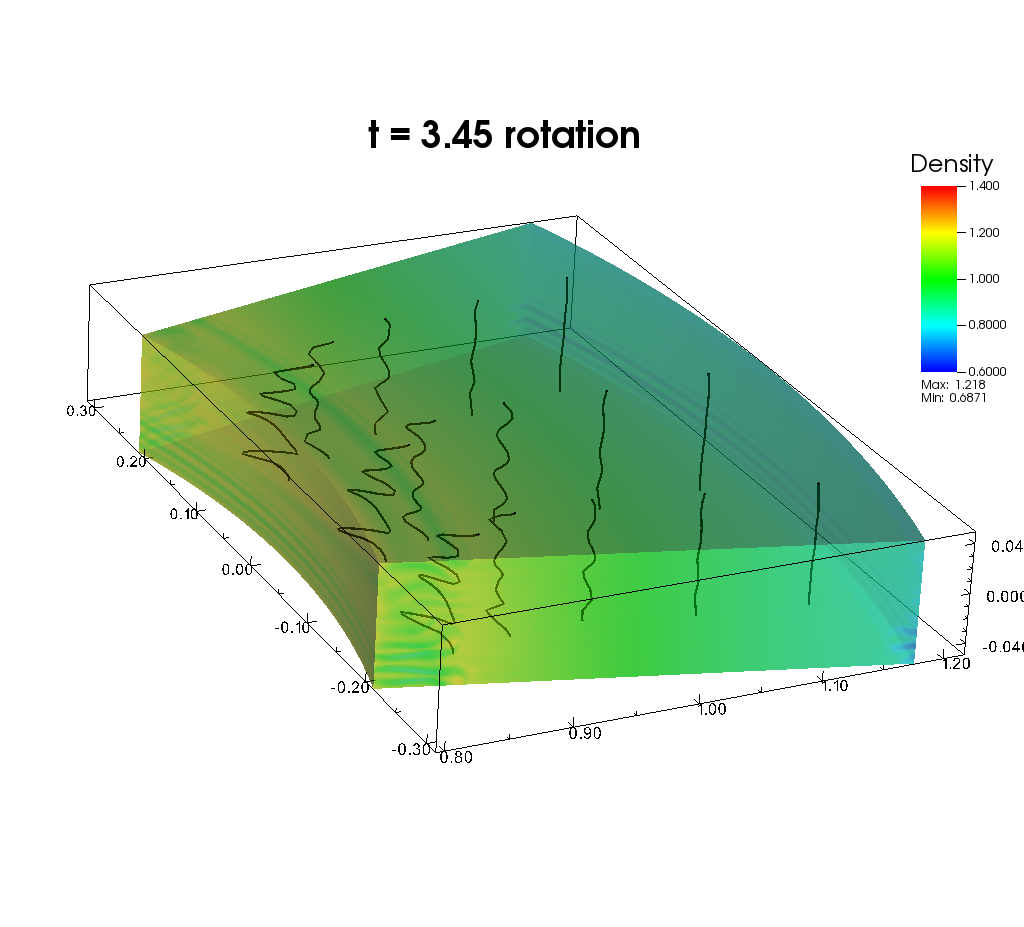}\\
    \vspace{-1.2cm}
    \includegraphics[width=0.45\textwidth]{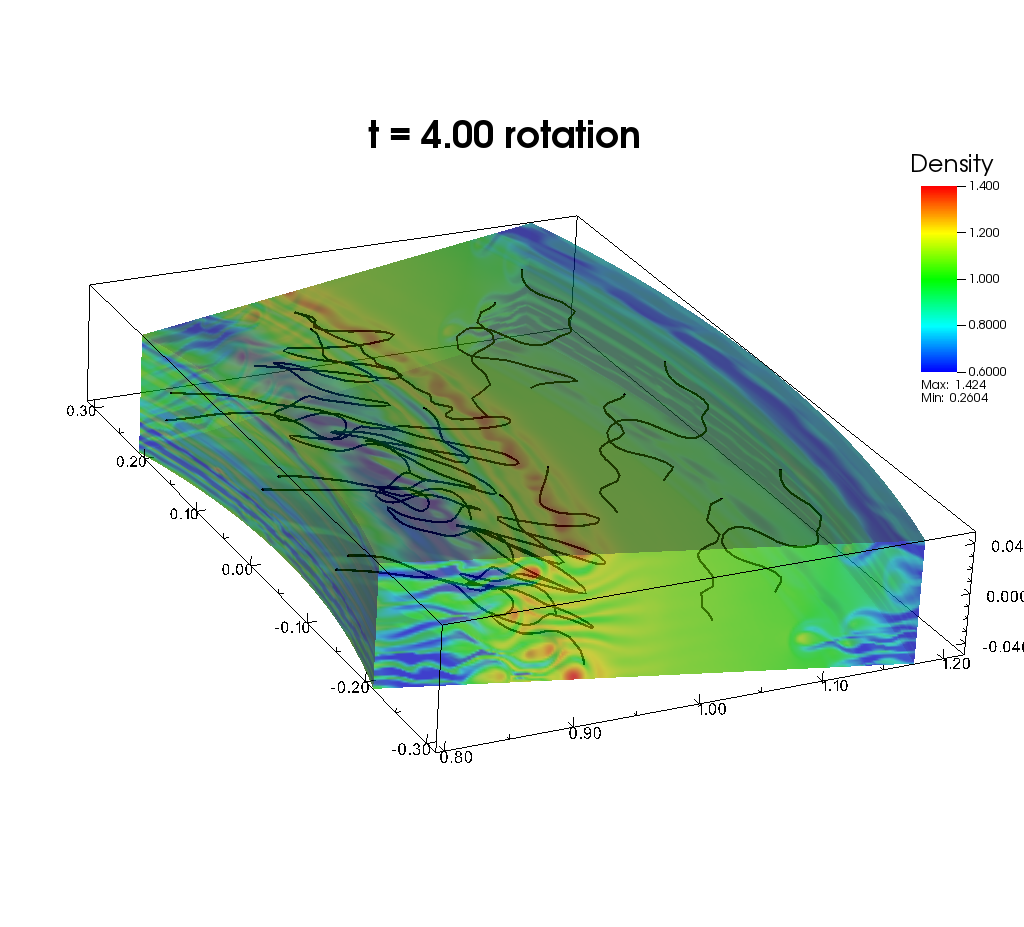}
    \includegraphics[width=0.45\textwidth]{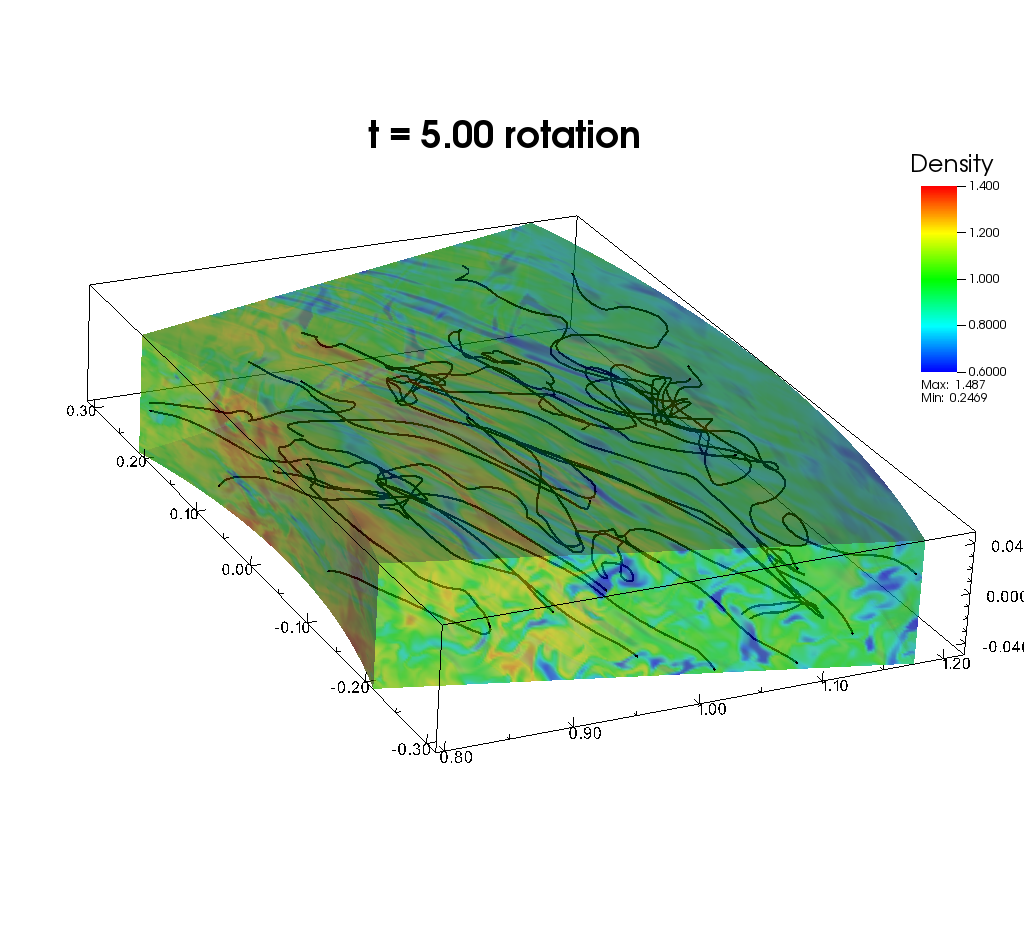}\\
    \vspace{-1.2cm}
    \includegraphics[width=0.45\textwidth]{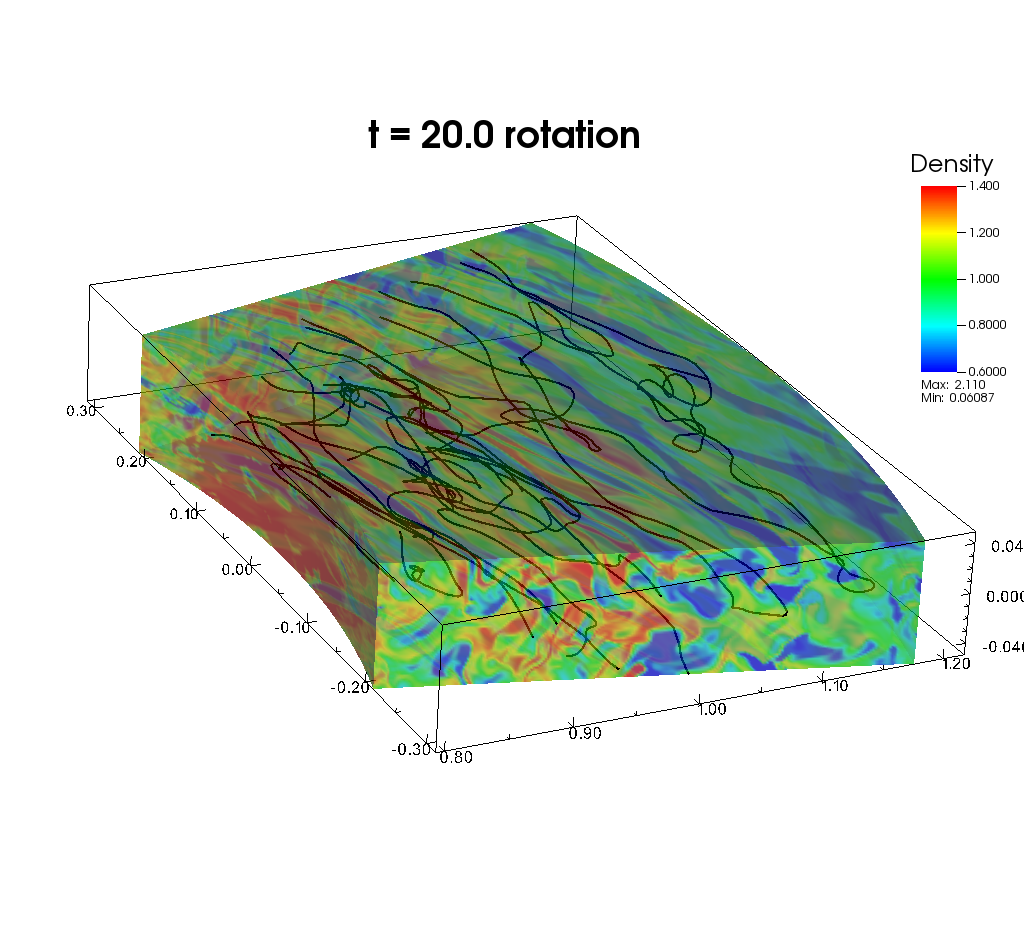}
    \includegraphics[width=0.45\textwidth]{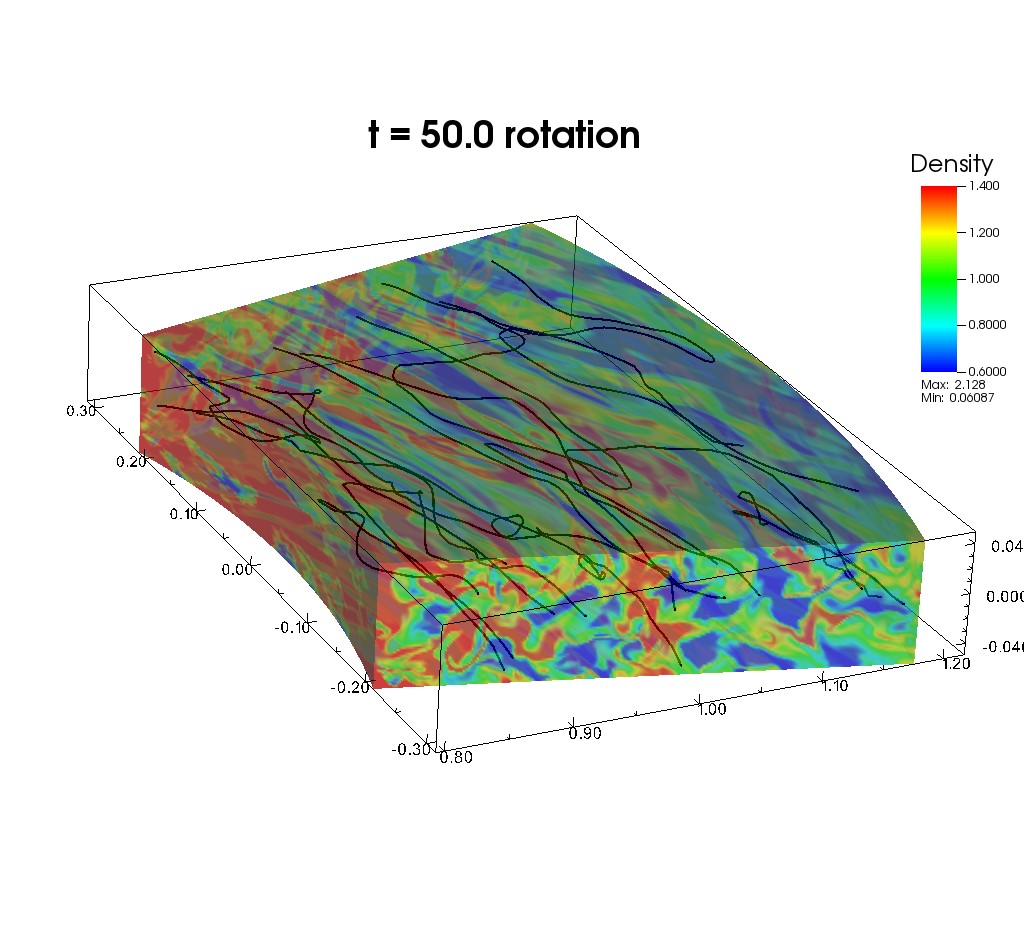}\\
    \vspace{-1.2cm}
    \includegraphics[width=0.45\textwidth]{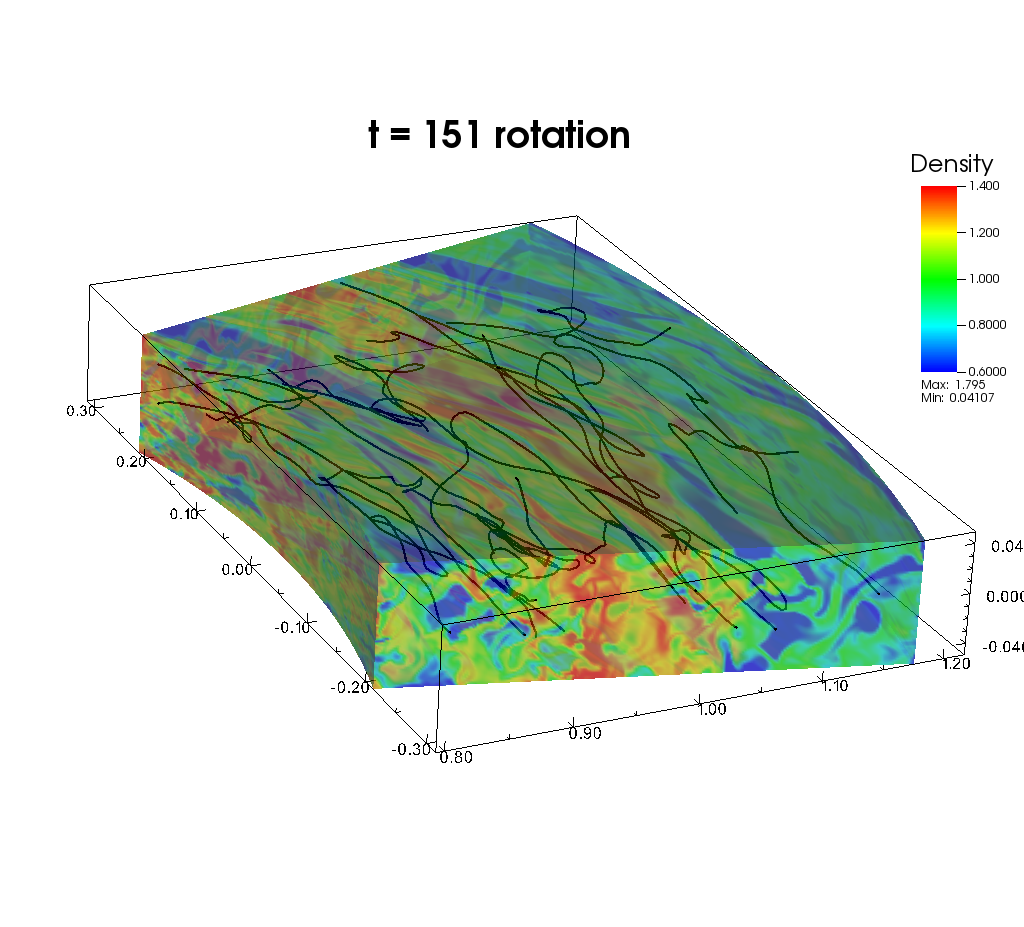}
    \includegraphics[width=0.45\textwidth]{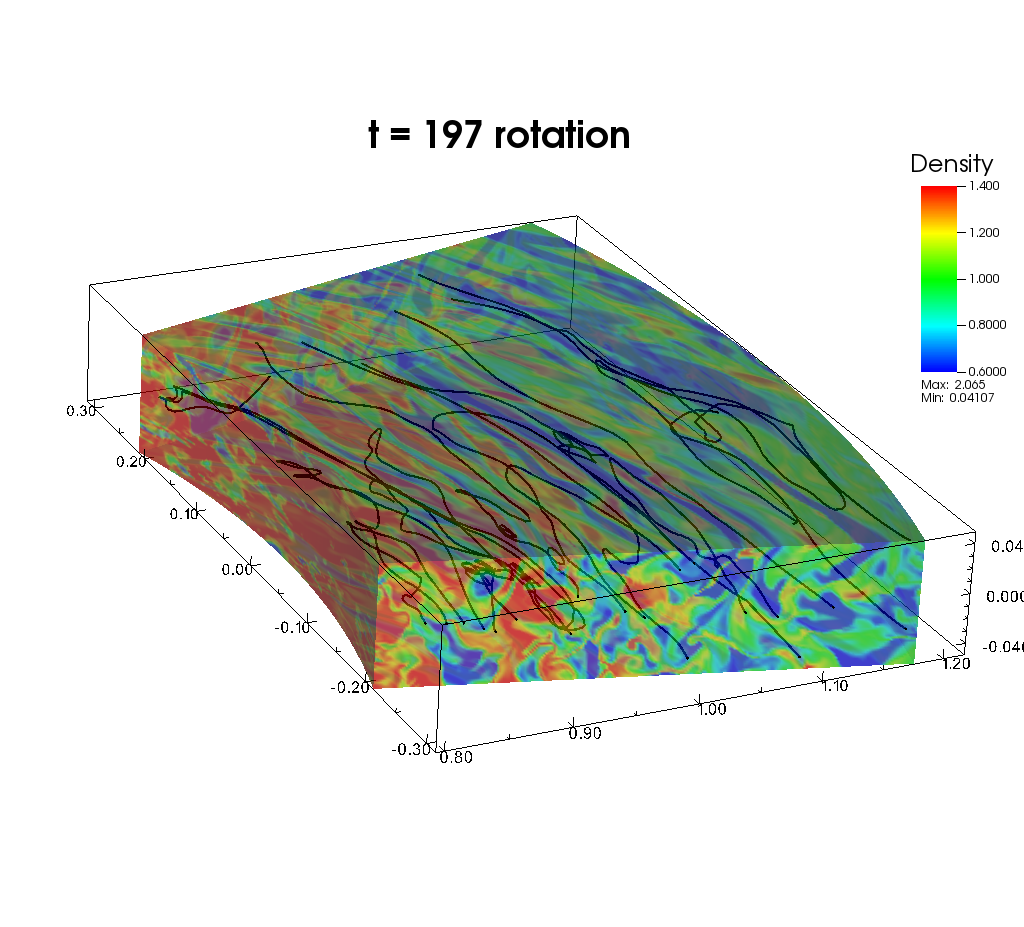}
  \end{center}
  \vspace{-1.5cm}
  \caption{Time evolution of the cylindrical case. Colors denote density and
    black lines indicate magnetic field lines. Movie is available at
    http://ea.c.u-tokyo.ac.jp/astro/Members/stakeru/research/cylshbx. 
    \label{fig:tevol30deg}}
\end{figure*}

\begin{figure*}
  \begin{center}
    \includegraphics[width=0.48\textwidth]{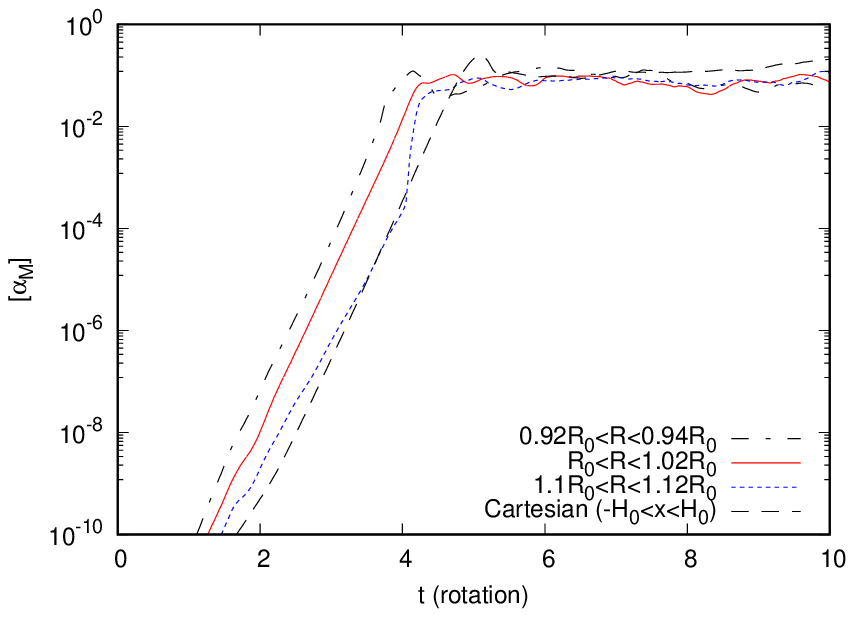}
    \includegraphics[width=0.48\textwidth]{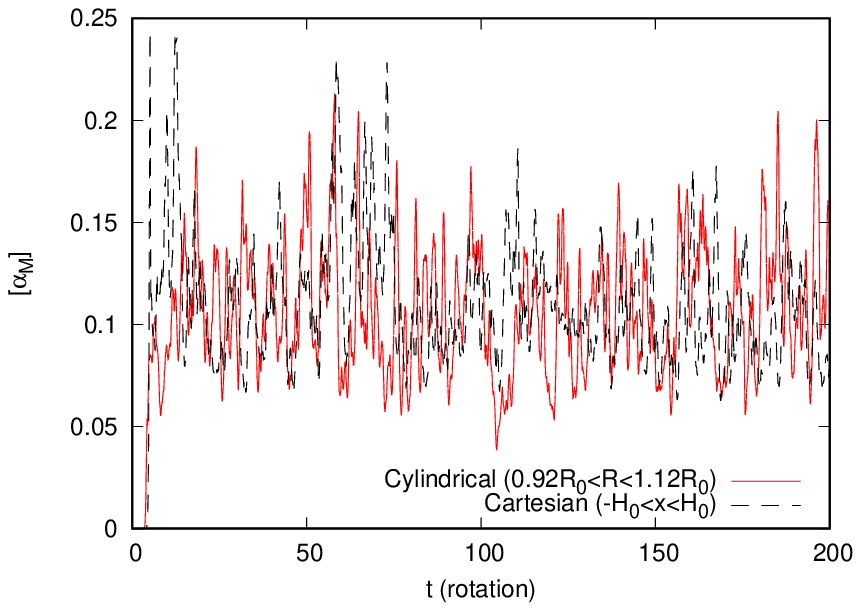}
    \caption{Time evolution of $\alpha_{\rm M}$. The left panel
      focuses on the early times of 
      $t<10(2\pi/\Omega_{\rm K,0})$. The black dot-dashed, red solid,
      and blue dotted lines respectively indicate the time evolution
      of inner ($0.92R_0<R<0.94R_0$), intermediate ($R_0<R<1.02R_0$),
      and outer ($1.1R_0<R<1.12R_0$) regions of the cylindrical
      shearing box.  The right panel presents the time
      evolution of a broader region of $0.92R_0 < R < 1.12R_0$ of the
      cylindrical shearing box by the red solid line until the end of
      the simulations at $t=200(2\pi/\Omega_{\rm K,0})$. 
      In both panels, the result of the Cartesian case averaged in the
      region of $-H_0 < x < H_0$ is also represented by black dashed
      lines for comparison. 
      \label{fig:alpha-t}
    }
  \end{center}
\end{figure*}

Figure \ref{fig:tevol30deg} shows 3D snapshots of the 
cylindrical case at eight different time slices.
The evolution at the early times ($=3-4$ rotations) exhibits that
the MRI starts to grow from inner locations, because the growth rate
is $\approx \frac{3}{4} \Omega_{\rm K} \propto R^{-3/2}$ \citep{bh91}
in this nearly Keplerian rotation condition. At $t=3.45$
rotations, the field lines in $R<R_0$ show channel-mode patterns, although
the outer field lines are still almost straight. At the slightly later times
at $t=4$ rotations, the inner region is already in the nonlinear regime, 
while the outer region is still in the linear growth stage of MRI.
Inspecting these two panels, one can also recognize the radial
dependence of $\lambda_{\rm max,init}(\propto R)$ in eq.(\ref{eq:lambdamax}).

The MRI initially excites radial magnetic field from the vertical field
as shown in these two panels. Later on the toroidal magnetic field is
amplified from $B_R$ by differential rotation. As a result, $B_{\phi}$
dominates the poloidal components at and after $t=5$ rotations. 
After $t\gtrsim 20$ rotations, the magnetic field is amplified to the
saturated state. 

The lower four panels show density perturbations are also excited in
the nonlinear saturation stage. At $t=50$ and 197 rotations, the
density fluctuations are larger than those at other times. At
$t=151$ rotations, a density bump is formed around $R\approx R_0$
(see also Figure \ref{fig:rhovph}),
although the overall density fluctuations in the entire box is
moderately smaller.

Figure \ref{fig:alpha-t} presents the time evolution of the
dimensionless volume averaged $R\phi$ component of the Maxwell stress,  
\begin{equation}
  [\alpha_{\rm M}]_{R_1}^{R_2}= \frac{-\int_{R_{1}}^{R_{2}}RdR\langle B_{R}B_{\phi}
    /4\pi\rangle }{\int_{R_{1}}^{R_{2}}RdR\langle \rho c_{\rm s}^2 \rangle},   
  \label{eq:aM}
\end{equation}
where from now on we define $\langle A \rangle$ as the $\phi$ and
$z$ integrated average of some variable, $A$, at $R$
\begin{equation}
  \langle A \rangle \equiv
  \frac{\int_{z_{-}}^{z_{+}}\int_{\phi_{-}}^{\phi_{+}} d\phi dz
    A}{\int_{z_{-}}^{z_{+}}\int_{\phi_{-}}^{\phi_{+}} d\phi dz}.
  \label{eq:pzave}
\end{equation}
By changing $R_1$ and $R_2$ in eq.(\ref{eq:aM}), we compare
$\alpha_{\rm M}$ in different regions; we set an inner region of
$0.92R_0 < R < 0.94R_0$, a middle region of $1R_0 < R < 1.02R_0$, and
an outer region of $1.1R_0 < R < 1.12R_0$.

The left panel shows the growth of $\alpha_{\rm M}$ in these three
different regions of the cylindrical shearing box at the early time
before 10 rotations, in comparison to the result of the Cartesian box.
The increase of $\alpha_{\rm M}$ is faster at smaller $R$, because the
growth rate of MRI is roughly proportional to $\Omega_{\rm K} \propto
R^{-3/2}$. The slope of the Cartesian case coincides with the slope of
the middle region ($R_0<R<1.02R_0$) of the cylindrical case, as
expected. However, the onset time of the Cartesian case is slightly
later. We do not know the exact reason of this time difference; it
is probably because of the effect of curvature \citep{lat15}. 

After $t\gtrsim 4$ rotations, the amplification of the magnetic
field almost saturates. The right panel compares the cylindrical and
Cartesian cases until the end of the simulation ($=200$ rotations).
For direct comparison, we picked regions with the same radial extent of
$2H_0$, $0.92R_0 < R < 1.12R_0$ and $-H_0 < x < H_0$, respectively.
Although the region of the radial box is $0.91R_0<R<1.11R_0$
if we choose the same grid number across $R=R_0$, we slightly shift it
outward to avoid the effect of the inner boundary (see Subsection
\ref{sec:zf}).  
While both cases exhibit intermittent behavior, the time averaged
values during $50-200(2\pi/\Omega_{\rm K,0})$ are quite similar;
the cylindrical case gives $\alpha_{\rm M}=0.106$ and the Cartesian
case gives $\alpha_{\rm M}=0.109$. 

\subsection{Radial Distribution}
We examine time, $\phi$, and $z$ averaged radial profiles of various
physical quantities in this subsection. The $\phi$ and $z$ averages
are taken by eq.(\ref{eq:pzave}). We take the time average from $t=50$
to 200 rotations, unless otherwise noted. 

\subsubsection{$\rho$ \& $v_{\phi}$}

\begin{figure}
  \begin{center}
    \includegraphics[width=0.45\textwidth]{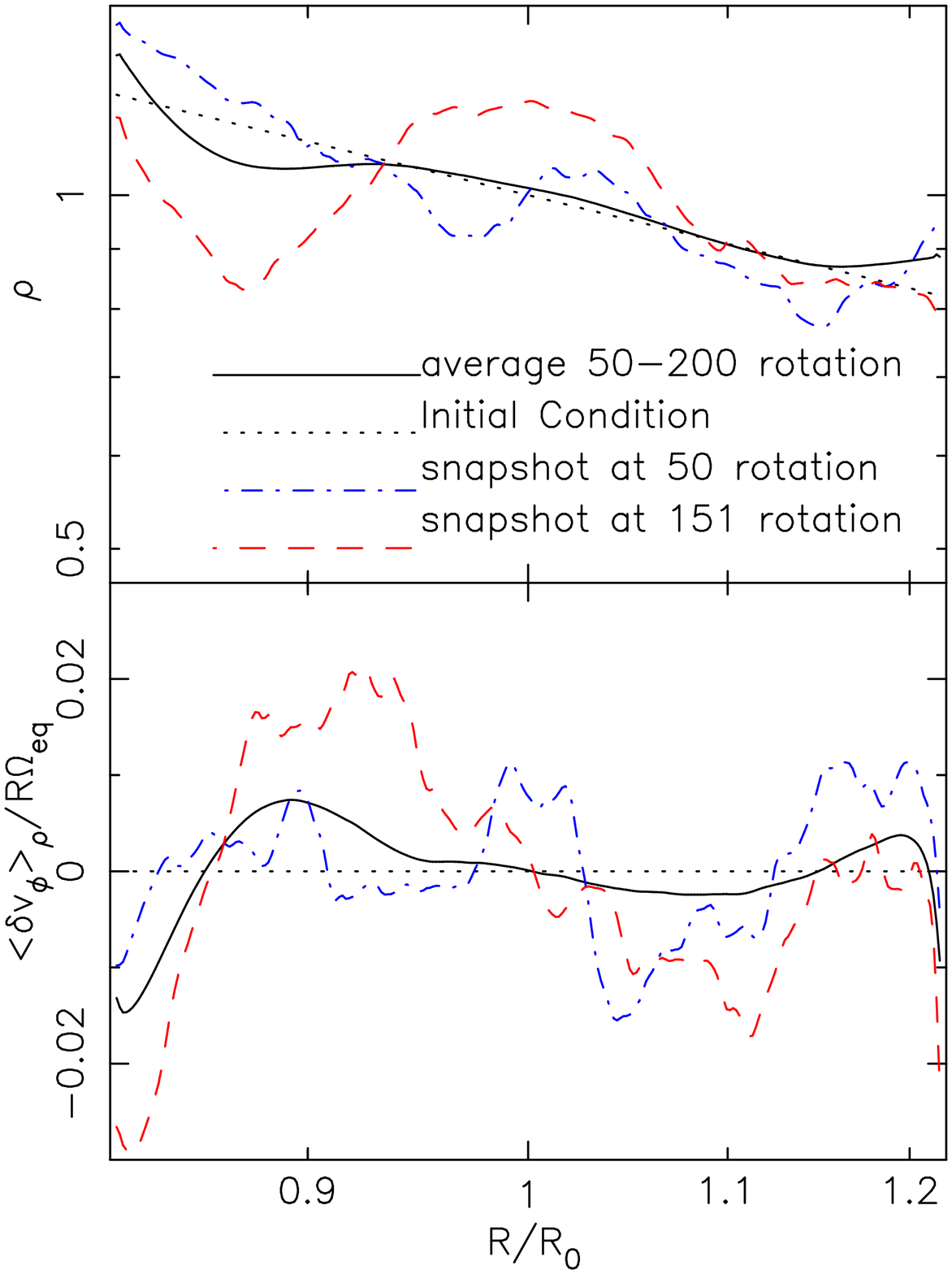}
    \caption{Comparison of radial profiles of $\phi$ and $z$ averaged
      density ({\it upper panel}) and dimensionless shift of azimuthal
      velocity from the initial equilibrium distribution 
      ($\delta v_{\phi}/R\Omega_{\rm eq}$; {\it lower panel})
      at different times. Black dotted and black solid lines
      indicate the initial condition and the time average from 50 --
      200 rotations, respectively. Blue dash-dotted and red dashed lines
      denote the snapshots at 50 rotations and 151 rotations,
      respectively, where the corresponding 3D snapshots are shown in
      Figure \ref{fig:tevol30deg}. 
      \label{fig:rhovph}
    }
  \end{center}
\end{figure}

The upper panel of Figure \ref{fig:rhovph} compares radial density
profiles at different times. The time-averaged distribution (black
solid line) shows that the initial profile (black dotted line) is
almost preserved. 
The deviations from the initial condition is larger near the inner and
outer boundaries. The gas slightly piles up near both boundaries
and the density there increases about 10\% from the initial value
because of boundary effects.

Snapshots at $t=50$ (blue dash dotted line) and 151 (red solid line)
rotations illustrates that the density distribution considerably
varies with time. At $t=151$ rotations, one can see a density bump
with the density enhancement by $\approx 15\%$ from the initial
condition near $R\approx R_0$, which can be also seen in the 3D
snapshot (Figure \ref{fig:tevol30deg}). This density enhancement is
a transiently formed zonal flow, which was also observed in Cartesian
shearing box simulations \citep[][]{joh09,sim18}.

The lower panel of Figure \ref{fig:rhovph} shows how the initial
equilibrium rotational profile is perturbed with time.
We present density weighted $\delta v_{\phi}$ (eq.\ref{eq:dvphi}),
\begin{equation}
  \langle\delta v_{\phi}\rangle_{\rho}
    \equiv \frac{\langle \rho \delta v_{\phi} \rangle}{\langle \rho
      \rangle},
\end{equation}
which is further normalized by the equilibrium rotational velocity
measured in the laboratory frame.

The time-averaged profile (solid line) shows that the dimensionless
$\delta v_{\phi}$ is kept small with $< 1.5\%$ in the entire region,
while the deviations are larger near the inner and outer boundaries
where the slope of $\langle \rho \rangle$ changes from the initial
condition. 
A steeper decrease of gas pressure with $R$ reduces the rotational
velocity near the inner boundary; a smaller 
contribution from the centrifugal force is sufficient to balance with
the inward gravity, because of the larger outward pressure gradient
force. Since in our simulation we assume the locally isothermal
condition, the pressure gradient force is modified only by the change
of a density gradient.  

Comparing the $\delta v_{\phi}$ and $\rho$ profiles, 
one can find that negative (positive) $\delta v_{\phi}$ corresponds to
the steeper (shallower) slope of density. While the two snapshots of
$\delta v_{\phi}$ roughly show the similar tendency, the detailed 
structures do not exactly follow it. This is because the radial force
balance is not always satisfied when the gas moves radially in a
time-dependent manner.

\subsubsection{Magnetic Field}
\label{sec:Bfield}
\begin{figure}
  \begin{center}
    \includegraphics[width=0.45\textwidth]{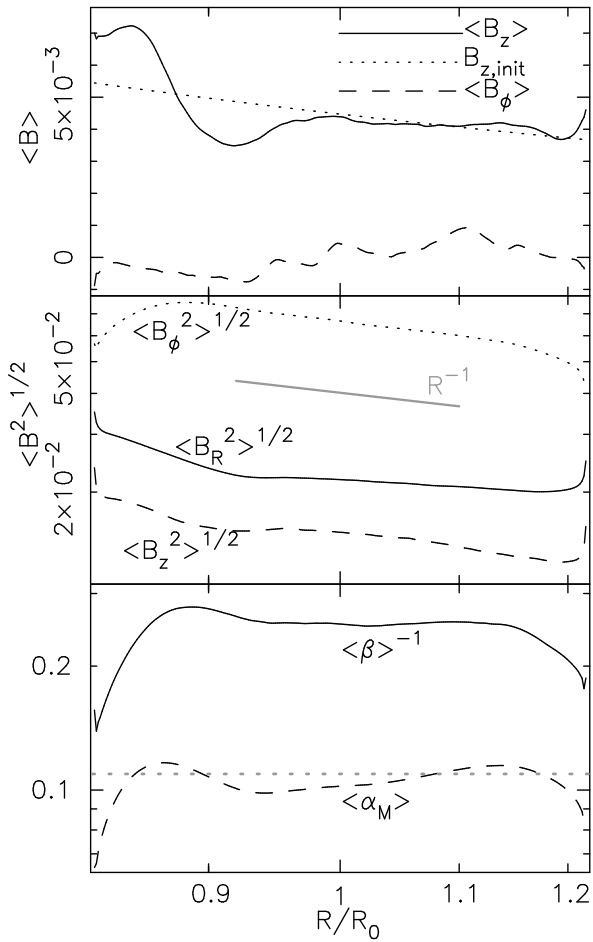}
    \caption{Radial distributions of various time-averaged quantities
      concerning the magnetic field. {\it Top}: The $z$ (solid) and $\phi$
      (dashed) components of the net magnetic flux density.
      The initial profile 
      of $B_z$ is also plotted for comparison. Note that $\langle
      B_R\rangle$ is not shown because it is strictly 0 by the
      conservation law of eq.(\ref{eq:phiphi}). {\it Middle}:
      Comparison of the $R$ (solid), $\phi$ (dotted), and $z$ (dashed)
      components of the root-mean-squared $B$.
      {\it Bottom}: The inverse of plasma $\langle\beta\rangle$ (solid) and the
      $R\phi$ component of the Maxwell stress, $\langle\alpha_{\rm M}\rangle$
      (dashed). The gray dotted lines are those from the $-H_0<x<H_0$
      region of the Cartesian shearing box. 
      \label{fig:Bfield}
   }
  \end{center}
\end{figure}

Figure \ref{fig:Bfield} presents various quantities of the magnetic
field. The top panel compares the radial profile of net magnetic
flux, $\langle B \rangle$, to the initial strength of the vertical
magnetic field. We note that $\langle B_R \rangle$ is kept to 0 within
the accuracy of a round-off error in our simulation because of 
the conservation law of 
eq.(\ref{eq:phiphi}),
and therefore we only present $\langle B_\phi\rangle$ (dashed)
and $\langle B_z \rangle$ (solid). 

This panel indicates that the initial profile of $B_z$ is roughly
preserved, although moderate pileups of $B_z$ are seen near both
boundaries, which are also formed by the influence of the radial
boundaries, as discussed in the density distribution (Figure
\ref{fig:rhovph}).  

$\langle B_{\phi}\rangle$ shows that the initial condition ($=0$) is
also almost conserved. We would like to note that the integrated
$\int_{R_{-}}^{R_{+}}dR R\langle B_{\phi} \rangle$ in the box is
strictly $0$ in our simulation by the conservation law of
eq.(\ref{eq:phiphi}) and the periodic condition at the $\phi$
boundaries. 

The middle panel of Figure \ref{fig:Bfield} presents the three
components of the root-mean-squared magnetic field, $\sqrt{\langle
  B^2\rangle }$, which is generally  much larger than $\langle
B\rangle$ by the contribution from the turbulent component.
The toroidal component dominates the poloidal ($R$ \& $z$) components
because the differential rotation winds up and amplifies $B_{\phi}$,
which is consistent with results obtained in local Cartesian shearing
box simulations \citep[e.g.,][]{hgb95,san04,dav10} and global simulations
\citep[e.g.,][]{arm98,haw00,si14}. 
The relative values in units of magnetic energy is
$B_{R}^2:B_{z}^2:B_{\phi}^2\approx 2:1:10-15$ except in the
regions near both boundaries.

The middle panel also shows that $\sqrt{\langle B_{\phi}^2\rangle}
\propto 1/R$ in $0.9R_0\lesssim R\lesssim 1.15R$. This trend is obtained
in previous global simulations \citep{flo11,si14}, which is anticipated
from the radial force balance between magnetic pressure and hoop stress,
\begin{equation}
  -\frac{1}{R^2}\frac{\partial}{\partial
    R}\left(R^2\frac{B_{\phi}^2}{8\pi}\right) = - \frac{\partial}{\partial
    R}\left(\frac{B_{\phi}^2}{8\pi}\right)  - \frac{B_{\phi}^2}{4\pi
    R} \approx 0.
\end{equation}
Near the radial boundaries, $\sqrt{\langle B_{\phi}^2\rangle}$ is
weaker than the strength expected from this trend. This is because the
differential rotation 
is weaker there, which corresponds to
$\frac{\partial \delta v_{\phi}}{\partial R}>0$ in Figure \ref{fig:rhovph},
and therefore the amplification of magnetic field is suppressed.
The poloidal components, which show a roughly similar radial dependence
to that of $B_{\phi}$, are basically controlled by the dominant toroidal
component, whereas they are also affected by the radial boundaries. 

The bottom panel of Figure \ref{fig:Bfield} presents $\alpha_{\rm M}$ 
(dashed line) and the inverse of a plasma $\beta$ value (solid line),
which is defined by the ratio of gas pressure to magnetic
pressure. Again, both quantities are averaged over the $\phi$
and $z$ components:
\begin{equation}
  \langle\alpha_{\rm M}\rangle= \frac{-\langle B_{R}B_{\phi}/4\pi\rangle}
  {\langle \rho c_{\rm s}^2 \rangle},   
  \label{eq:aM2}
\end{equation}
and
\begin{equation}
  \langle\beta\rangle^{-1} \equiv \frac{(\langle B_R^2 + B_{\phi}^2 + B_z^2)
    /8\pi\rangle}{\langle \rho c_{\rm s}^2\rangle}.
  \label{eq:beta}
\end{equation}

The results of the Cartesian shearing box are also plotted (gray
dotted lines) for comparison. Both $\langle\beta\rangle^{-1}$ and
$\langle\alpha_{\rm M}\rangle$
show almost flat dependence on $R$ except in the regions near the
boundaries, and their values also agree with those of the Cartesian
case within 10\% difference. $\langle\beta\rangle^{-1}\approx 0.25$ in the flat
region ($0.93R_0<R<1.15R_0$), which indicates that the magnetic energy
($\propto B^2$) is amplified by 250 times from the initial condition,
$\beta_{z,{\rm init}}^{-1}=10^{-3}$ (eq.\ref{eq:btzint}). 

The magnetic pressure, which is dominated by
$B_{\phi}^2/8\pi$, is proportional to $R^{-2}$, as shown in the middle
panel. This dependence is the same as that of the gas
pressure, since $\rho\propto R^{-1}$ is adopted at the radial shearing
periodic boundaries (eq.\ref{eq:smplsc1}) and $c_{\rm s}^2\propto
R^{-1}$ is fixed in our locally isothermal assumption. Therefore, the
$R^{-2}$ dependences of both numerator and denominator of
eq.(\ref{eq:beta}) are canceled out so that $\langle\beta\rangle^{-1}$
is nearly a constant on $R$.

$\langle \alpha_{\rm M}\rangle$ is also a nearly constant but
slightly increases with $R$ in the middle region that is not affected
by the boundaries. This weak dependence is important in the transport
of angular momentum and consequent mass accretion, which is discussed
in the next subsection.

\subsubsection{Angular Momentum \& Accretion}
\label{sec:angmomaccr}

\begin{figure*}
  \begin{center}
    \includegraphics[width=0.32\textwidth]{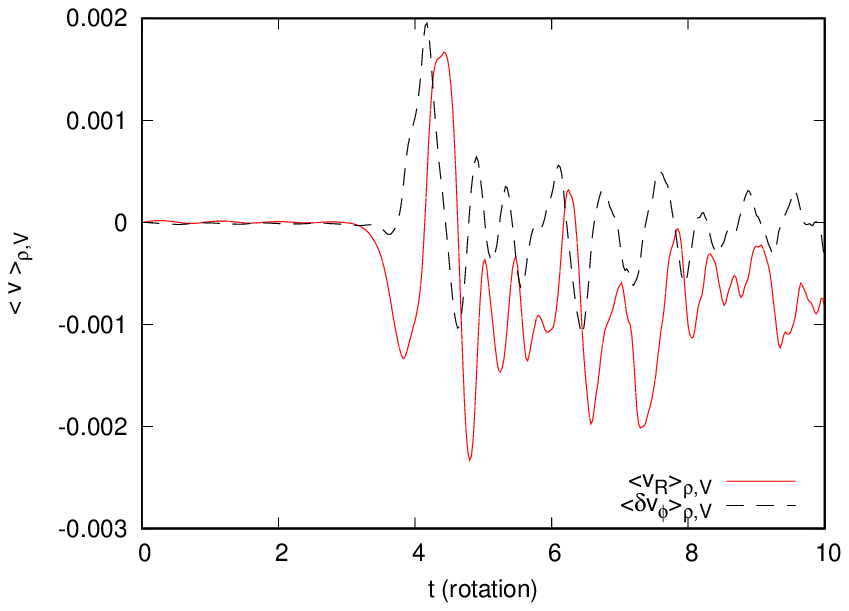}
    \includegraphics[width=0.32\textwidth]{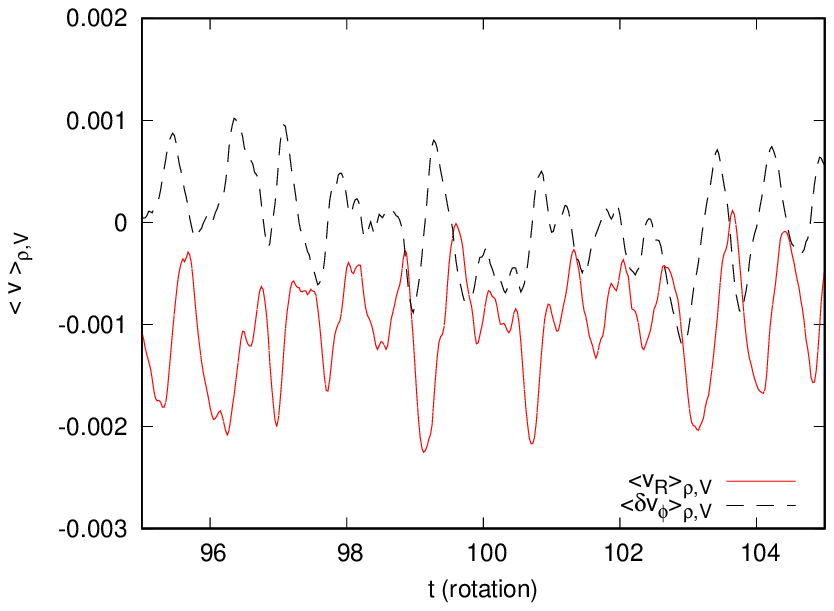}
    \includegraphics[width=0.32\textwidth]{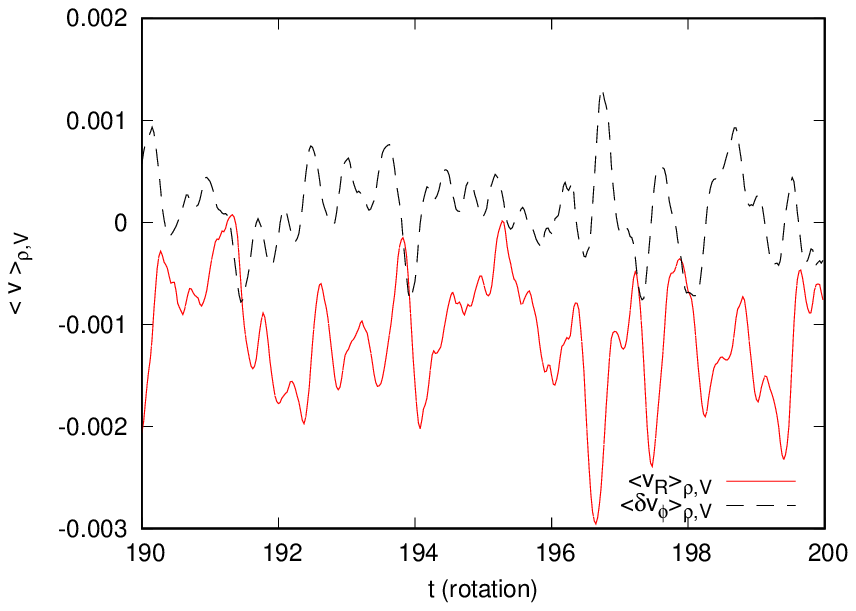} 
    \caption{Time evolution of the density weighted and volume averaged $v_R$
      (solid red lines) and $\delta v_{\phi}$ (black dashed lines).
      The left, middle, and right panels respectively show the initial,
      intermediate, and final 10 rotations of the simulation. 
      \label{fig:vrvphtev}
    }
  \end{center}
\end{figure*}

\begin{figure}
  \begin{center}
    \includegraphics[width=0.45\textwidth]{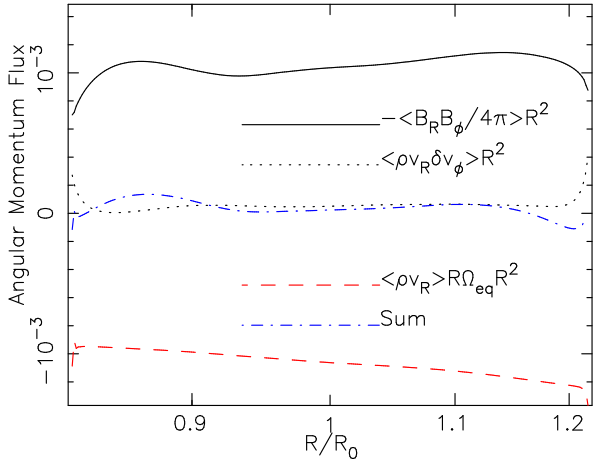}
    \caption{Comparison of different components of time-averaged
      angular momentum fluxes. The black solid, black dotted, and red
      dashed lines respectively show the angular momentum fluxes
      carried by the $R\phi$ component of the Maxwell stress, the turbulent
      Reynolds stress, and the net mass accretion. The blue dash-dotted
      line represents the sum of these three components.
      \label{fig:angflx}
    }
  \end{center}
\end{figure}
A great advantage of our cylindrical shearing box approach to the
Cartesian shearing box setup is that we can handle radial mass
accretion directly. In order to realize this, we do not impose a
shearing periodic constraint on the total angular momentum at the
radial boundaries but instead constrain the turbulent part (see
eq.(\ref{eq:Aangmom}) in Subsection \ref{sec:rbc} \& Appendix
\ref{sec:numrsb}).   
The total angular momentum in the simulation box is not conserved, and
mass accretion or decretion can be automatically induced by the loss
or gain of angular momentum. 
In other words, we liberate the center of mass in the box from a
fixed origin and test whether time-steady mass accretion is actually
achieved by the outward transport of angular momentum via excited MHD
turbulence. 

Let us examine the time evolution of the radial and angular momentums in
the simulation box.   
Figure \ref{fig:vrvphtev} presents the density weighted and volume averaged
horizontal velocities, $\langle v_{R} \rangle_{\rho, V}$ and
$\langle \delta v_{\phi} \rangle_{\rho, V}$, where $\langle v \rangle_{\rho, V}
\equiv [\rho v]_{_V} /[\rho]_{_V}$. From eqs.(\ref{eq:epiR}) and
(\ref{eq:epiphi}) we can derive the solutions that represent epicyclic
oscillations with an arbitrary velocity amplitude, $a$: 
\begin{equation}
  \langle\delta v_{\phi}\rangle_{\rho,V} \approx a\sin\left(\Omega_{\rm eq,0}t
  + \delta \right)
  \label{eq:episolphi}
\end{equation}
\begin{equation}
  \langle v_{R}\rangle_{\rho,V} \approx 2a\sin\left(\Omega_{\rm eq,0}t
  + \delta -\frac{\pi}{2}\right),   
  \label{eq:eqisolR}
\end{equation}
where $\delta$ is a phase shift. These solutions show that the phase of
$\langle v_{R}\rangle_{\rho,V}$ is delayed by $\pi/2$ 
from that of $\langle\delta v_{\phi}\rangle_{\rho,V}$ and the amplitude
of $\langle v_{R}\rangle_{\rho,V}$ is twice that of
$\langle\delta v_{\phi}\rangle_{\rho,V}$

Readers can recognize that the oscillatory behavior of the horizontal
velocities in Figure \ref{fig:vrvphtev} roughly follow the characteristics
of these epicyclic oscillations,
although it is considerably perturbed from time to time by the magnetic field
and the curvature effects that are not considered in the solutions of
eqs. (\ref{eq:episolphi}) and (\ref{eq:eqisolR}).  The left panel of
Figure \ref{fig:vrvphtev} shows that the simulation box starts to oscillate
at $t\gtrsim 3$ rotations when the magnetic field is amplified by the MRI.
While $\langle \delta v_{\phi} \rangle_{\rho,V}$ oscillates around 0,
the center of the oscillation of $\langle v_{R}\rangle_{\rho,V}$ slowly shifts
downward; the mass accretion is gradually induced.

In the middle (95-105 rotations) and right (190-200 rotations) panels,
$\langle v_{R}\rangle_{\rho,V}$ does not decrease monotonically but it
oscillates roughly around $\approx 0.001$. This indicates that the
mass accretion occurs in a quasi-steady manner, if we take a time average
covering the duration of $\gtrsim 10$ rotations. 
On the other hand, the oscillation of $\langle \delta v_{\phi}
\rangle_{\rho,V}$ is still kept around $\approx 0$ at later times.
This clearly shows that the total angular momentum is almost conserved
for the long-time average. We can conclude that, by utilizing the shearing
variable of the angular momentum, $S_{\rm mom,\phi}$, (eq.\ref{eq:Aangmom}),
the time-steady mass accretion can be realized while keeping
the angular momentum in the box conserved, as we aimed in Subsection
\ref{sec:rbc}.  

Next, we inspect the radial profile of different components of angular momentum
fluxes when the mass accretes in a quasi-steady manner. 
Taking the $\phi$ and $z$ integrated average under the periodic
boundary condition and assuming the steady-state condition,
$\partial_t \cdots = 0$, we can obtain an equation that 
describes the balance of angular momentum fluxes in the laboratory frame 
( see eq.\ref{eq:angmom}) as
\begin{equation}
  \frac{\partial}{\partial R}\left[R^2\left(\langle \rho v_R\rangle
    R\Omega_{\rm eq} + \langle \rho v_R\delta v_{\phi}\rangle -
    \frac{\langle B_R B_{\phi}\rangle}{4\pi}\right)\right] = 0, 
  \label{eq:angmomblnc}
\end{equation}
where the first term indicates the angular momentum flux carried by 
net radial flows, the second term is that by the turbulent Reynolds
stress, and the third term is that by the Maxwell stress. We note that in
the Cartesian shearing box approach, mass accretion rate, $\dot{M}$,
is estimated from the second and third terms by using this equation,
\begin{equation}
  \dot{M} \equiv - 2\pi \int dz R \rho v_{R} \approx
  -\frac{2\pi}{\Omega_{\rm K}} \int dz \left\langle \rho v_R\delta v_{\phi}
  - \frac{B_R B_{\phi}}{4\pi}\right\rangle, 
\end{equation}
even though net $\langle v_R \rangle=0$. Our cylindrical shearing box
can directly test the justification of this conventional approach. 

Figure \ref{fig:angflx} compares the three terms of eq.(\ref{eq:angmomblnc}),
where the time averages are again taken from 50 to 200 rotations.   
The outward transport of angular momentum is mainly done by the Maxwell
stress (black solid). The turbulent Reynolds stress (black dotted)
also transports angular momentum outward, however its
contribution is $\lesssim 1/10$ times smaller than that from the
Maxwell stress in most of the simulation region. 
On the other hand, the sign of the accretion term is negative, which
indicates that the angular momentum is carried inward by the net mass
accretion.

The sum of these three terms (blue dash-dotted line) is
nearly 0; the balance between the outward transport by the MHD
turbulence and the inward transport by the mass accretion is almost
satisfied, and the total angular momentum is conserved in a
self-regulating manner after the magnetic field is amplified to the
saturated state, even though we do not impose a constraint on the total
angular momentum.   

From the conservation law of eq.(\ref{eq:Amass}), our simulation
gives $\langle \rho v_R R\rangle =$ const. We adopt the nearly
Keplerian rotational velocity for the equilibrium state, which roughly
gives $\Omega_{\rm eq} \propto R^{-3/2}$. These relations leads to the $R$
scaling of the first term of eq.(\ref{eq:angmomblnc}) as
$R^2\langle\rho v_{R}\rangle R \Omega_{\rm eq}\propto R^{1/2}$. The
weak radial dependence of the accretion term in Figure
\ref{fig:angflx} reflects this $R^{1/2}$ scaling. The Maxwell stress
also shows the same dependence of $-\langle B_R B_{\phi}/4\pi\rangle R^2
\propto R^{1/2}$
to balance with the accretion term.
This dependence is consistent with $\langle\alpha_{\rm
  M}\rangle\propto R^{1/2}$ in Figure \ref{fig:Bfield},  
and further implies that the (dimensional) turbulent viscosity,
$\nu_{\rm M}$, of the Maxwell stress has the relation of 
$\nu_{\rm M}\approx \langle \alpha_{\rm M}\rangle c_{\rm s}H \approx \langle
\alpha_{\rm M}\rangle c_{\rm s}^2/\Omega_{K}\propto R$.  

\subsubsection{Mass Accretion and Radial Transport of $B_z$}
\begin{figure}
  \begin{center}
    \includegraphics[width=0.45\textwidth]{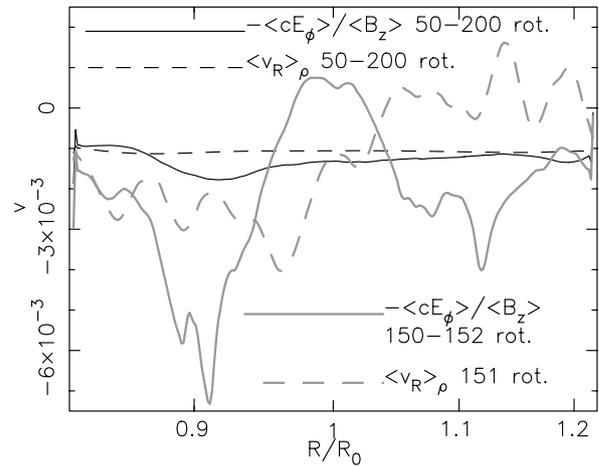}
    \caption{Comparison of the radial flows of gas ($\langle
      v_R\rangle_{\rho}$; dashed lines)  and vertical magnetic field
      ($\langle v_R\rangle_{B_z} = \langle cE_{\phi}\rangle/ \langle
      B_z \rangle$; solid lines). The thin 
      black lines are the time average from 50-200 rotations, and the gray
      lines are the snapshots at 151 rotations, where we take the average
      of $\pm$ one rotations for $\langle v_R\rangle_{B_z}$. 
      \label{fig:vrBz}
    }
  \end{center}
\end{figure}
In the previous subsection we have discussed the mass accretion from
a viewpoint of the angular momentum balance. In this subsection, we further
inspect radial flows of not only mass but also vertical magnetic field.
As discussed in \citet{si14}, the radial velocity of the gas and that of 
the vertical magnetic fields do not generally coincide, even if the ideal
MHD condition is considered, because of the turbulent diffusion of
magnetic fields. The radial flow of $B_z$ determines the pileup or
diffusion of the poloidal magnetic field in a disk, and consequently
controls the long-time evolution of the large-scale magnetic field
\citep{lub94,rl08,go12,oku14,to14}.  

The radial velocity of gas is taken from the density weighted average, 
\begin{equation}
  \langle v_R \rangle_{\rho} = \frac{\langle \rho v_R \rangle}
          {\langle \rho \rangle}, 
\end{equation}
where the subscript $\rho$ is put to explicitly show gas flow.
For the radial velocity of $B_z$, we introduce  
\begin{equation}
  \langle v_R \rangle_{B_z} = -\frac{\langle cE_{\phi}\rangle}{\langle B_z\rangle} 
  = \frac{\langle v_R B_z - v_z B_R\rangle}{\langle B_z \rangle},
  \label{eq:vRBz}
\end{equation}
which is expected from the $z$ component of the induction equation
(eq.\ref{eq:ind}). Taking the $\phi$ and $z$ integration of the equation that
describes the time variation of $B_z$, we get
\begin{equation}
  \frac{\partial \langle B_z\rangle}{\partial t} + \frac{1}{R}
  \frac{\partial}{\partial R} \left[R(-\langle cE_{\phi}\rangle )\right] = 0. 
  \label{eq:indzEp}
\end{equation}
The form of eq.(\ref{eq:indzEp}) is essentially an equation of
continuity for $\langle B_z \rangle$, and therefore, we use
eq.(\ref{eq:vRBz}) to follow the radial motion of $\langle B_z\rangle$.   

Figure \ref{fig:vrBz} compares $\langle v_{R}\rangle_{\rho}$ (dashed) and
$\langle v_{R}\rangle_{B_z}$ (solid). The time averaged gas flow (thin black
dashed line) shows the gas accretes inward with a constant
$\langle v_{R}\rangle_{\rho}$, which is consistent
with eq.(\ref{eq:smplsc2}) and the discussion in Sub-subsection
\ref{sec:angmomaccr}.  

The time-averaged radial velocity of the vertical magnetic field also shows
a nearly constant $\langle v_{R}\rangle_{B_z}$. However, the inward velocity
is slightly faster, $\langle v_{R}\rangle_{B_z} < \langle v_{R}\rangle_{\rho}
(< 0)$, in most of the region except near the inner boundary, which indicates
that the vertical magnetic flux drifts inward through the gas.
As discussed above, the magnetic field is not strictly frozen into the
gas, even though the ideal MHD condition is imposed on the
simulation.
The inward velocity of $B_z$ is decelerated from $R=0.9R_0$ to $0.85R_0$, 
which leads to the pileup of the vertical magnetic flux, as discussed
in Figure \ref{fig:Bfield}.

The snapshot profiles of $\langle v_{R}\rangle_{\rho}$ and
$\langle v_{R}\rangle_{B_z}$ at $t=151$ rotations are also plotted in
Figure \ref{fig:vrBz}. As for $\langle v_{R}\rangle_{B_z}$, we take the average
from $t=150$ to 152 rotations, because
the pure snapshot gives spuriously huge values at locations where $\langle
B_z \rangle$ is occasionally $\approx 0$.

Both $\langle v_{R}\rangle_{\rho}$ and $\langle v_{R}\rangle_{B_z}$ indicate
that the mass accretion and the inward transport of magnetic flux do not
occur in a time-steady manner. The direction of the gas flow is inward in
the inner side ($R\lesssim R_0$) and outward in the outer side ($R\gtrsim R_0$)
at $t=151$ rotations because diverging flows are excited from the density
bump that is formed at this time (Figures \ref{fig:tevol30deg} \&
\ref{fig:rhovph}). 

The snapshot of $\langle v_{R}\rangle_{B_z}$ (gray solid line) largely deviates
from that of $\langle v_{R}\rangle_{\rho}$ (gray dashed line); the radial
motion of $B_z$ drifts from the accreting gas because of turbulent diffusion and
reconnection \citep{lv99}. 

\section{Discussion}
\label{sec:discuss}

\subsection{Treatment of Radial Boundaries}
\label{sec:bddis}
After the simulation starts from the initial condition of
$v_R=\mbf{0}$, mass accretion is gradually induced by the excited MHD
turbulence that transports angular momentum outward
(Figure \ref{fig:vrvphtev}). We did not impose any constraint on the
mass accretion rate or the angular momentum transport rate.  
The mass accretion rate is determined by the balance between the
angular momentum fluxes from mass accretion and MHD turbulence
in a self-consistent and self-regulating manner. 
Each component of the time-averaged radial angular momentum flux
shows a smooth and monotonic profile in $R$ (Figure \ref{fig:angflx}). 
The treatment of the radial boundary condition works well
at least for handling the time-averaged properties of the mass accretion. 

However, there are issues concerning the boundary treatment that
should be addressed in future work. The first point arises from the
difference between $\Omega_{\rm eq,+}$ and $\Omega_{\rm eq,-}$ at the
$R_{\pm}$ boundaries. At the linear stage of the MRI, the magnetic
field grows first at the inner boundary because the growth time
($\propto \Omega_{\rm eq}$) is shortest there. A part of the amplified
magnetic field at the inner boundary is transported to the outer
boundary and into the simulation domain because of the shearing periodic 
condition (Figure \ref{fig:tevol30deg}), which does not occur in
realistic accretion disks.   
For this reason, we have to be careful when we focus on specific
phenomena, such as individual channel flows, near the radial
boundaries.  On the other hand, we expect that the radial boundary
treatment gives reasonable time-averaged properties, provided that
the appropriate shearing variables, $S$, are adopted (Subsection
\ref{sec:rbc}).  

Another possible concern is the propagation of waves across the radial
boundary, which is also related to the difference between $\Omega_{\rm
  eq,+}$ and $\Omega_{\rm   eq,-}$. We can expand the basic MHD
equations (eqs.\ref{eq:mass}--\ref{eq:divB}) into the mean and
fluctuating components, and MHD waves are derived from the 
latter component. The current formulation using the shearing variables
focuses only on the mean component and does not take special care of
the fluctuating component. Therefore, waves that propagate
across the radial boundary could suffer partial reflection.
The treatment of the fluctuating component should be done as a next
step.


\subsection{Zonal Flows}
\label{sec:zf}

The radial distribution of the density exhibits bumps and dips.
Although the amplitudes of the radial density variations are not so
large, they are not erased even for the time average over 50 -- 200
rotations (Figure \ref{fig:rhovph}). Because of the bumpy profile of
the density and, accordingly, the pressure, the azimuthal velocity
also deviates from the equilibrium value with $\delta v_{\phi}<0$
($>0$) in regions with a steeper (shallower) density gradient than the
equilibrium gradient. As a result, the differential rotation is not
constant in $R$. 
The toroidal magnetic field is more amplified in the regions with stronger
differential rotation, $\frac{\partial \delta v_{\phi}}{\partial R} < 0$.
As a result, the unsigned toroidal magnetic field,
$\sqrt{B_{\phi}^2}$, is not a monotonically decreasing function of $R$ but  
shows a peak at $R\approx 0.88R_{0}$, as discussed in Sub-subsection
\ref{sec:Bfield}. 

Although the bumpy density structures, or zonal flows, are created
physically \citep{joh09}, they may be affected by the radial 
boundaries because the deviation of $\langle\delta v_{\phi}\rangle$
from 0 is larger near both radial boundaries. In particular, it is  
more prominent near the inner boundary because the curvature effect
($\propto 1/R$) is more severe there.   
The partial reflection of propagating waves at the
radial boundary (Subsection \ref{sec:bddis}) may cause the bumpy
density structure. 

In addition, the numerical implementation may cause the bumpy density
structure. We adopt the CT scheme to
update the magnetic field. The locations of the three components of
the magnetic field are different from that of the other physical
variables. Therefore, interpolation is required to use the shearing
variables with magnetic fields, which causes truncation errors and
numerical diffusion.  
Our specific implementation method is described in Appendix
\ref{sec:numrsb}. Although we carefully chose the interpolation method
after much trial and error, it still may not be a perfect one. More
elaborate and innovative methods will be explored in future work.

\subsection{Radial Dependences and Shearing Variables}
In this paper we presented one simulation with a single set of radial
dependences for the density, the temperature, and the vertical
magnetic field strength.
The power-law index of the density, $q_{\rho}=1$, is required from the
shearing conditions for $S_{\rm mass}$ (eq.\ref{eq:Amass})
and $S_{{\rm mom},r}$ (eq.\ref{eq:rmom}).
The power-law index of the temperature, $q_T=1$, is chosen to give $c_{\rm s}
\propto R^{-1/2}$, which is the same scaling as that of the equilibrium
rotation velocity. The power-law index of the initial net vertical magnetic
flux, $q_{B}=1$, is regulated from the adopted $q_{\rho}$ and $q_{T}$ to
give a constant $\beta_{z,{\rm init}}$. 

In general, however, the radial dependences are determined
independently of each other. 
Therefore, it is worth pursuing cases with different sets of power-law
indices to study various types of accretion disks, which we plan to tackle
in our future studies. 

Among the three power-law indices, $q_{\rho}$ needs to be treated carefully.
The adopted $q_{\rho}=1$ is consistent with the conservation of mass via
$S_{\rm mass}$ (eq.\ref{eq:Amass}) and radial momentum via $S_{{\rm mom},r}$
(eq.\ref{eq:Armom}). When a different $q_{\rho}$ is adopted, we cannot satisfy
the shearing variables of both $S_{\rm mass}$ and $S_{{\rm mom},r}$
simultaneously, and have to dismiss either one of then.

It is better to keep $S_{\rm mass}$ rather than $S_{{\rm mom},r}$, because
even in the present formulation the radial momentum is conserved only
in an approximate sense (Subsection \ref{sec:rbc}). However, in this
case the radial dynamical pressures, $(\rho v_R^2 R)_{\pm}$, at the
$R_{\pm}$ boundaries are not balanced, and hence, the simulation box
will be accelerated to the $+$ or $-R$ direction. A prescription to
prevent this systematic acceleration must take into account the
magnetic terms (see eq.\ref{eq:rmom}) in $S_{{\rm mom},r}$.

\subsection{Future Applications}
Although there is room to improve the treatment of the
radial shearing boundary (Subsection \ref{sec:bddis}), the
cylindrical shearing box model has various possible extensions and 
applications. 

\subsubsection{Vertical Stratification}
A first extension of the cylindrical shearing box framework takes into
account the stratification of density by the vertical component of the
gravity of a central object. 

In recent years, vertical outflows and disk winds have been widely
discussed that they play a significant role in the evolution of
protoplanetary disks \citep{fer06,suz16,tm18}, and they are studied
in vertically stratified Cartesian shearing box simulations
\citep{si09,suz10,bai13,bs13a,les13,fro13,rio16,mor19}.
The magnetic centrifugal force often plays an important role in
driving disk winds \citep{bp82}.
In addition to MHD turbulence, magnetocentrifugal acceleration that
removes angular momentum from a disk
causes the accretion of gas 
\citep{pp92}.

In principle it is quite difficult, and probably impossible, to properly
treat the magnetocentrifugal acceleration with the Cartesian shearing
box model because of the $\pm x$ symmetry (Section \ref{sec:intro}). 
The vertical component of the angular momentum flux is evaluated from the
$yz$ component of the Maxwell and Reynolds stresses in Cartesian
coordinates. However, the sign of the vertical angular momentum flux
is ambiguous because of the $\pm x$ symmetry; 
it is flipped when the central object is switched from the $-x$
direction to the $+x$ direction.

In contrast, there is no such ambiguity in the sign of the angular momentum
flux in the cylindrical approach. 
The cylindrical shearing box with vertical stratification can properly
evaluate the removal rate of angular momentum by magnetocentrifugal
driven disk winds.  

There are some issues that are not present in the Cartesian shearing
box when we include the vertical density stratification in the
cylindrical shearing box. The first issue is the radial dependence of
the scale height. For example, we presented the case with $H\propto
R$, which is derived from $q_T=1$. When we apply the radial shearing
boundary conditions to a vertically stratified box, the radial
dependence of $H$ needs to be taken into account in a consistent way.

Another point is that the equilibrium rotational velocity generally involves
vertical shear \citep[see, e.g.][]{si14}. The gravity of a central object
is weaker at higher altitudes. Therefore, rotational velocities are
usually slower at higher altitudes
for the same $R$, though this can be reversed by the contribution from
the pressure gradient force. We note that there is an attempt to
consider the vertical shear in the Cartesian shearing box by \citet{mp15}. 

\subsubsection{Spherical Coordinates}
We can extend our framework of the cylindrical shearing box to
spherical coordinates in a straightforward manner. When the vertical
stratification is taken into account, it is probably better to adopt
spherical coordinates rather than cylindrical coordinates, as in the
``spherical disks'' by \citet{kla03}, because the disk scale height 
usually increases with distance from the origin.

\subsubsection{Physical Processes}
In the presented simulation we solved the ideal MHD equations with a
locally isothermal equation of state, which is the simplest setting
for demonstrative purposes. It is possible to consider various
physical processes in the cylindrical shearing box as is done in
Cartesian shearing box simulations.

For example, self-gravity can be included in the momentum equation to
study the formation of stars, brown dwarfs, and planets
\citep{gam01,hs19}. 
To determine realistic temperature distributions in various types of
disks, radiative cooling and heating should be considered in the
energy equation \citep[][]{tur03,hir06,shi10,jia13}. 
If the temperature is not high and the ionization is not sufficient,
as expected in protoplanetary disks, the magnetic diffusion by
non-ideal MHD effects 
needs to be taken into account \citep{ss02,bs13b,kun13,mp17}. 

\subsubsection{Particles}
The shearing box model is also a strong tool to study the dynamics of
particles in accretion disks.

The energization of non-thermal particles in accretion disks around
compact objects has been investigated by particle-in-cell
simulations in Cartesian shearing boxes \citep[e.g.,][]{hos15,kun16}.
One of the severe problems of using the Cartesian box is the
existence of unphysical runaway particles; once the gyroradius of a
particle exceeds the radial box size, it continuously gains the energy
as a result of acceleration \citep{kim16}. Therefore, we cannot
determine the maximum energy of the accelerated particles in the
Cartesian shearing box model. 
In reality, however, the acceleration eventually saturates when the
gyroradius becomes comparable to the size of the system
\citep{kim19}. The cylindrical shearing box approach can handle the
saturation of the energy gain because it includes the curvature; the
size of the acceleration region is regulated by the curvature radius.

The Cartesian shearing box approach is often adopted to study the
dynamics of dust grains in protoplanetary disks
\citep[e.g.,][]{car06,gre12,zhu15}. The pressure gradient force 
induces the inward drift of dust grains from the background gas
\citep{ada76}. While this radial drift can be taken into account in
the Cartesian shearing box model as an external force \citep{joh06},
the cylindrical shearing box can consider it in a self-consistent
way, which can be a reliable method to understand reasonable pathways
for the planet formation \citep[e.g.,][]{kob16}. 

The Cartesian shearing box model also considers larger bodies in
protoplanetary disks, such as planetesimals and (proto)planets
\citep{np04,yan09,mut10,tan12}.
One of the targets of this type of simulations is to understand the
migration of (proto)planets. The direction and rate of the migration
are primarily determined by the difference between the torques exerted
by density waves excited from the inner and outer locations of the 
planet \citep{tan02,cm07,bar14,kan18}. In addition, they are also
affected by the radial flow of the background gas \citep{ogi17}.
It is quite difficult to quantitatively and directly determine the
small difference between the inner and outer torques from the
background gas flow in the Cartesian
shearing box mainly because of the symmetry with respect to
the $\pm x$ directions. In contrast, our cylindrical approach would be
a powerful tool to solve this problem.

\section{Summary}
\label{sec:sum}
We developed the basic framework of the cylindrical shearing box,
focusing on MHD simulations for accretion disks. We constructed the
shearing periodic boundary conditions at the radial boundaries by 
utilizing the conservation relations of the basic MHD equations. While
the cylindrical shearing box is basically a local approach, it also
takes into account global effects from the curvature of cylindrical
coordinates.  
One of the great advantages of our treatment is that we can directly
capture the net mass accretion, which cannot be handled by the
Cartesian shearing box treatment because of the 
radial symmetry.

We performed the MHD simulation in the unstratified cylindrical
shearing box with a moderate resolution that resolves one scale height
by 64 grid points. 
Inward mass flows are naturally induced by the outward flux of angular
momentum carried by the MHD turbulence. While the local cylindrical
simulation box oscillates quasi-periodically as a result of the
epicyclic motion, the total angular momentum averaged over $\gtrsim
10$ rotations is conserved by the balance between the 
inward angular momentum flux advected by the accreting mass and the
outward angular momentum flux by the MHD turbulence. The quasi-time-steady
accretion is realized in our cylindrical shearing box simulation.
The basic physical properties of the excited MHD turbulence, such as
the saturation level of the amplified magnetic fields, are similar to
those obtained from the Cartesian shearing box.  

While the global effects of curvature are considered, the cylindrical
shearing box framework still has the advantage of the local approach
that (i) fine-scale phenomena of the turbulence can be resolved by
zooming in on a local patch of the accretion disk and (ii) long-time
simulations can be performed stably within an acceptable
computational time. Related to the point (ii), it took only $\sim$
a day for the presented case with a medium resolution of 64 grids
per $H_0$ (Table \ref{tab:cyl}) to run up to 200 rotations on a
standard parallel computer with 512 CPU cores. It would be possible to
perform simulations with a similar resolution up to several
thousand rotations within a realistic computational time. This could
be quite an efficient tool to study long-time evolution governed by
the timescale of diffusion.  

It is still not easy to run global simulations for long times
($\sim 10^{3-4}$ dynamical timescales).
Global simulations usually cover a large dynamic range from a fast
rotating inner region to a slow rotating outer region
\citep[e.g.,][]{flo11,si14}. Therefore, in order to follow several
thousand rotations at the region of interest, usually located at an
intermediate region in the simulation domain, it is necessary to 
cover larger rotation times at the inner region, which is not
realistic with the current computational resources.

There is still room to improve the numerical implementation of the
radial shearing boundary condition, in particular for the treatment of
propagating waves.
As discussed in Section \ref{sec:discuss}, the cylindrical shearing
box framework has various applications, which
are open to future works by all those who are interested. 

Numerical computations were carried out on Cray XC40 at YITP, Kyoto
University, and Cray XC50 at Center for Computational Astrophysics,
National Astronomical Observatory of Japan.
We thank Geoffroy Lesur, James Stone, and Charles Gammie
for valuable and critical comments on an earlier version of the draft. 
The authors thank the referee for many constructive comments.
This work was supported by Grants-in-Aid for Scientific Research from
the MEXT of Japan, 17H01105.

\onecolumn
\begin{appendix}
\section{Treatment of External Forces}
\label{sec:extforce}
The radial component of the momentum equation, eq.(\ref{eq:mom}), is written as
\begin{displaymath}
  \frac{\partial v_R}{\partial t} + v_R\frac{\partial v_R}{\partial R}
  + \frac{v_{\phi}}{R}\frac{\partial v_R}{\partial \phi}
  + v_z\frac{\partial v_R}{\partial z} - \frac{v_{\phi}^2}{R}
  = - \frac{1}{\rho}\frac{\partial}{\partial R}\left(p
  + \frac{B_{\phi}^2 + B_z^2}{8\pi}\right)
\end{displaymath}
\begin{equation}
  + \frac{B_{\phi}}{4\pi\rho R}
  \frac{\partial B_R}{\partial \phi} + \frac{B_z}{4\pi\rho}
  \frac{\partial B_R}{\partial z} - \frac{B_{\phi}^2}{4\pi\rho R}
  -\frac{GM_{\star}}{R^2} + R \Omega_{\rm eq,0}^2 + 2\Omega_{\rm eq,0}v_{\phi}. 
\label{eq:rdmom}
\end{equation}
The mutual subtraction of the external forces and the curvature term
($v_{\phi}^2/R$) causes the numerical cancellation of significant digits.
Therefore, it is better to consider the deviation from the equilibrium profile. 
In the equilibrium state, the radial force balance 
\begin{equation}
  F_{\rm eq} \equiv \frac{v_{\phi,{\rm eq}}^2}{R} - \frac{1}{\rho_{\rm eq}}
  \frac{\partial p_{\rm eq}}{\partial R} - \frac{GM_{\star}}{R^2}
  + R \Omega_{\rm eq,0} ^2+ 2\Omega_{\rm eq,0}v_{\phi,{\rm eq}} = 0
  \label{eq:rdfc}
\end{equation}
is satisfied, where
\begin{equation}
  - \frac{1}{\rho_{\rm eq}}\frac{\partial p_{\rm eq}}{\partial R}
  = (q_{\rho} + q_T)\frac{c_{\rm s,0}^2}{R}\left(\frac{R}{R_0}\right)^{-q_T}.
\end{equation} 
Substituting eq.(\ref{eq:rdfc}) into eq.(\ref{eq:rdmom}), we obtain
\begin{displaymath}
  \frac{\partial v_R}{\partial t} + v_R\frac{\partial v_R}{\partial R}
  + \frac{v_{\phi}}{R}\frac{\partial v_R}{\partial \phi}
  + v_z\frac{\partial v_R}{\partial z} 
  = - \frac{1}{\rho}\frac{\partial}{\partial R}\left(p
  + \frac{B_{\phi}^2 + B_z^2}{8\pi}\right)
\end{displaymath}
\begin{equation}
  + \frac{B_{\phi}}{4\pi\rho R}
  \frac{\partial B_R}{\partial \phi} + \frac{B_z}{4\pi\rho}
  \frac{\partial B_R}{\partial z} - \frac{B_{\phi}^2}{4\pi\rho R}
  + 2\left(\Omega_{\rm eq,0} + \frac{v_{\phi,{\rm eq}}}{R}\right)\delta v_{\phi}
  + \frac{\delta v_{\phi}^2}{R}
  - (q_{\rho} + q_T)\frac{c_{\rm s,0}^2}{R}\left(\frac{R}{R_0}\right)^{-q_T}. 
  \label{eq:exterm}
\end{equation}  
We use this expression with $\delta v_{\phi}$ for updating $v_R$ to reduce
numerical errors.

\section{Formulae}
\subsection{Basic Equations in the Rest Frame}
\label{sec:beq}
We summarize basic equations in conservative forms in the rest frame.
The $R$-derivatives of the following equations are used for the shearing
variables presented in Subsection \ref{sec:rbc}.  

The mass conservation is expressed as
\begin{equation}
  \frac{\partial \rho}{\partial t} + \frac{1}{R}\frac{\partial}{\partial R}
  (\rho u_R R) + \frac{1}{R}\frac{\partial}{\partial \phi}(\rho u_{\phi})
  + \frac{\partial}{\partial z}(\rho u_z)= 0. 
  \label{eq:massap}
\end{equation}
The radial component of momentum flux evolves as 
\begin{displaymath}
  \frac{\partial}{\partial t}(\rho u_R) + \frac{1}{R}\frac{\partial}{\partial R}
  (\rho u_R^2 R) + \frac{1}{R}\frac{\partial}{\partial \phi}(\rho u_R u_{\phi})
  + \frac{\partial}{\partial z}(\rho u_R u_z) = \rho \frac{u_{\phi}^2}{R}
  - \frac{\partial p}{\partial R} - \rho \frac{GM_{\star}}{R^2}
\end{displaymath}
\begin{equation}  
  + \frac{1}{R^2}\frac{\partial}{\partial R}\left(\frac{B_R^2R^2}{8\pi}\right)
  - \frac{1}{R^2}\frac{\partial}{\partial R}\left(\frac{B_{\phi}^2R^2}{8\pi}\right)
  - \frac{\partial}{\partial R}\left(\frac{B_z^2}{8\pi}\right) + \frac{1}{4\pi}
  \left[\frac{1}{R}\frac{\partial}{\partial \phi}(B_R B_{\phi})
    + \frac{\partial}{\partial z}(B_R B_z)\right], 
  \label{eq:rmom}
\end{equation}
where the gravity and a curvature term ($u_{\phi}^2/R$) need to be treated
as source terms. These two terms and the gas pressure gradient term constitute
the main part of radial force balance. Numerical treatment of these terms
is described in Appendix \ref{sec:extforce}. 

The evolution of angular momentum flux is
\begin{displaymath}
  \frac{\partial}{\partial t}(\rho u_{\phi} R) + \frac{1}{R}
  \frac{\partial}{\partial R}(\rho u_R u_{\phi}R^2) + \frac{1}{R}
  \frac{\partial}{\partial \phi}(\rho u_{\phi}^2 R)
  + \frac{\partial}{\partial z} (\rho u_{\phi} u_z R) = -\frac{1}{R}
  \frac{\partial}{\partial \phi}(p R)
\end{displaymath}
\begin{equation}
  - \frac{1}{R}\frac{\partial}{\partial \phi}\left[(B_R^2 + B_z^2)R\right]
  + \frac{1}{R}\frac{\partial}{\partial \phi}
  \left(\frac{B_{\phi}^2R}{8\pi}\right) + \frac{1}{4\pi R}
  \frac{\partial}{\partial R}(B_R B_{\phi} R^2)
  + \frac{1}{4\pi}\frac{\partial}{\partial z}(B_{\phi} B_z R^2).
  \label{eq:angmom}
\end{equation}
The vertical component of momentum flux evolves as
\begin{displaymath}
  \frac{\partial}{\partial t}(\rho u_z) + \frac{1}{R}\frac{\partial}{\partial R}
  (\rho u_R u_z R) + \frac{1}{R}\frac{\partial}{\partial \phi}(\rho u_{\phi}u_z)
  + \frac{\partial}{\partial z}(\rho u_z^2) = -\frac{\partial}{\partial z}
  \left(p + \frac{B_R^2 + B_{\phi}^2}{8\pi}\right)
\end{displaymath}
\begin{equation}  
  + \frac{\partial}{\partial z}
  \left(\frac{B_z^2}{8\pi}\right) + \frac{1}{4\pi R}\frac{\partial}{\partial R}
  (B_R B_z R) + \frac{1}{4\pi R} \frac{\partial}{\partial \phi}(B_{\phi}B_z).
  \label{eq:zmom}
\end{equation}

The three components of the induction equation (eq.\ref{eq:ind}) are
\begin{equation}
  \frac{\partial B_R}{\partial t} = \frac{1}{R}\frac{\partial}{\partial \phi}
  (u_R B_{\phi} - u_{\phi}B_R) - \frac{\partial}{\partial z}(u_z B_R - u_R B_z),
  \label{eq:indR}
\end{equation}
\begin{equation}
  \frac{\partial B_{\phi}}{\partial t} = \frac{\partial}{\partial z}
  (u_{\phi}B_z - u_z B_{\phi}) - \frac{\partial}{\partial R}
  (u_R B_{\phi} - u_{\phi}B_R),
  \label{eq:indphi}
\end{equation}
and
\begin{equation}
  \frac{\partial B_z}{\partial t} = \frac{1}{R}\frac{\partial}{\partial R}
       [R(u_z B_R - u_R B_z)] - \frac{1}{R}\frac{\partial}{\partial \phi}
       (u_{\phi}B_z - u_z B_{\phi}),
       \label{eq:indz}
\end{equation}
respectively. These evolutionary equations are constrained by
\begin{equation}
  \frac{1}{R}\frac{\partial}{\partial R}(RB_R) + \frac{1}{R}
  \frac{\partial B_{\phi}}{\partial \phi} + \frac{\partial B_z}{\partial z}
  = 0. 
\end{equation}

The total energy equation can be written in a conservative form:
\begin{displaymath}
  \frac{\partial}{\partial t}\left(\frac{1}{2}\rho u^2 + \rho e
  + \frac{B^2}{8\pi}\right) + \frac{1}{R}\frac{\partial}{\partial R}
  \left[R\left\{\rho u_R\left(\frac{u^2}{2} + e + \frac{p}{\rho}\right)
    + u_RB^2 - B_R(\mbf{u\cdot B})\right\}\right]
\end{displaymath}
\begin{equation}
  + \frac{1}{R}
  \frac{\partial}{\partial \phi}\left[\rho u_{\phi}\left(\frac{u^2}{2}
    + e + \frac{p}{\rho}\right)+ u_{\phi}B^2 - B_{\phi}(\mbf{u\cdot
      B})\right] +\frac{\partial}{\partial z}\left[\rho u_{z}
    \left(\frac{u^2}{2} + e + \frac{p}{\rho}\right)
    + u_{z}B^2 - B_{z}(\mbf{u\cdot B})\right] = 0.
  \label{eq:toteng}
\end{equation}

\subsection{Transformation between the Rest and Corotating Frames}
\label{sec:corrst}
The equations in the rest frame shown in the previous subsection
can be easily derived by replacing $\mbf{v}$ by $\mbf{u}$ and removing
the inertial terms of eqs.(\ref{eq:mass}) -- (\ref{eq:divB}) in
the corotating frame. When we transform from one frame to the other frame,
for example, to deal with orbital advection \citep{fargo16}, 
we have to keep in mind that the meanings of the time derivatives are
different in these two frames. 
Below we show the transformation of the $\phi$ component of the induction
equation between the rest and corotating frames for a representative example;
other equations can be derived in a similar manner. 

The $R$ and $z$ derivatives of the terms with $u_{\phi}$ in
eq.(\ref{eq:indphi}) can be expressed by $v_{\phi}$ as 
\begin{equation}
  \frac{\partial}{\partial R}(u_{\phi}B_R) = \Omega_{\rm eq,0}
  \frac{\partial}{\partial R} (R B_R) + \frac{\partial}{\partial R}
  (v_{\phi}B_R)
  \label{eq:rstcor1}
\end{equation}
and
\begin{equation}
  \frac{\partial}{\partial z}(u_{\phi}B_z) = R\Omega_{\rm eq,0}
  \frac{\partial B_z}{\partial z} + \frac{\partial}{\partial
    z}(v_{\phi}B_z).
  \label{eq:rstcor2}
\end{equation}
From $\mbf{\nabla\cdot B}=0$, we obtain 
\begin{equation}
  \Omega_{\rm eq,0}\frac{\partial}{\partial R}(R B_R) + R\Omega_{\rm
    eq,0} \frac{\partial B_z}{\partial z} = - \Omega_{\rm eq,0}
  \frac{\partial B_{\phi}}{\partial \phi}
  \label{eq:OmgdivB}
\end{equation}
Substituting eqs.(\ref{eq:rstcor1})--(\ref{eq:OmgdivB}) into
eq.(\ref{eq:indphi}), we get
\begin{equation}
  \left(\frac{\partial B_{\phi}}{\partial t}\right)_{\rm corot}
  = \left(\frac{\partial B_{\phi}}{\partial t}\right)_{\rm rest} +
  \frac{R\Omega_{\rm eq,0}}{R}\frac{\partial B_{\phi}}{\partial \phi}
  = \frac{\partial}{\partial z} (v_{\phi}B_z - v_z B_{\phi}) -
  \frac{\partial}{\partial R} (v_R B_{\phi} - v_{\phi}B_R),  
\end{equation}
where subscripts, ``corot'' and ``rest'' are the Eulerian time
derivatives in the corotating frame and in the rest frame,
respectively.

\section{Numerical Treatment of Radial Shearing Boundary}
\label{sec:numrsb}
\subsection{Basic Concept}
\begin{figure}
  \begin{center}
    \includegraphics[width=0.9\textwidth]{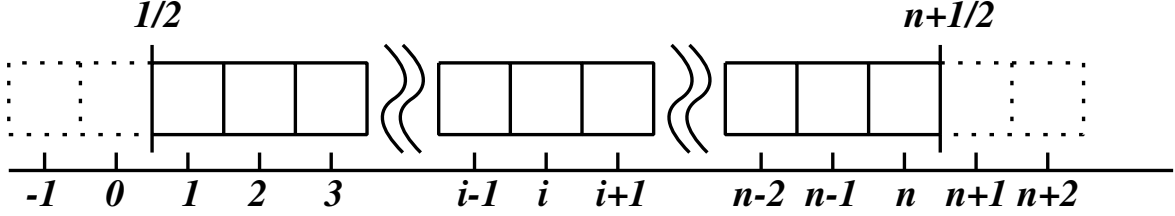}
  \end{center}
  \caption{
    Labels for radial grid points. The simulation region is covered by
    the cells from $i=1$ to $n$, namely the inner edge is located
    at $i=1/2$ and the outer edge is at $i=n+1/2$. The cells shown by dashed
    lines indicate ghost cells. 
    \label{fig:radmesh}
  }
\end{figure}

Before describing our specific method, we summarize the basic
concept of the numerical treatment for the radial shearing boundary. 
Let us consider a simulation box that is covered by $n$ grid points
from $i=1$ to $i=n$ along the $R$ axis. We set inner ghost
cells at the grid points of $i=0, -1, \cdots$ and outer ghost cells at
$i=n+1, n+2, \cdots$ (Figure \ref{fig:radmesh}). The exact simulation
region is from the $i=1/2$ boundary between the first active cell
($i=1$) and the inner neighboring ghost cell ($i=0$) to the $i=n+1/2$
boundary between the $n$th active cell ($i=n$) and the outer
neighboring ghost cell ($i=n+1$). 


If we pick out the time derivative terms and the radial derivative terms
of eqs.(\ref{eq:massap}) -- (\ref{eq:zmom}) and (\ref{eq:indphi}) --
(\ref{eq:toteng}), we can write the corresponding finite difference equation
in a symbolic form as follows: 
\begin{equation}
  \frac{V^{m+1}_{i}-V^{m}_{i}}{\Delta t} + \frac{F^{m+\nicefrac{1}{2}}_{i+\nicefrac{1}{2}}
    - F^{m+\nicefrac{1}{2}}_{i-\nicefrac{1}{2}}}{\xi_{R}} = 0, 
  \label{eq:fdeq}
\end{equation}
where the superscripts indicate labels for time and the subscripts correspond
to radial locations; $\xi_{R}=\Delta (\frac{1}{2}R^2) = R\Delta R$ except
for eq.(\ref{eq:indphi}) where $\xi_R = \Delta R$. 
Eq. (\ref{eq:fdeq}) updates $V_i$ from $t=m$ to $t=m+1$ with 2nd order
accuracy in time.

The shearing variables, $S$, are derived directly from the flux, $F$,
in eq.(\ref{eq:fdeq}), whereas we neglected terms with small
contributions in Subsection \ref{sec:rbc}. 
A direct numerical implementation of the shearing boundary condition is
to impose
\begin{equation}
  S^{m+\nicefrac{1}{2}}_{\nicefrac{1}{2},j_{-}} = S^{m+\nicefrac{1}{2}}_{n+\nicefrac{1}{2},j_{+}}
  \label{eq:Ardsh}
\end{equation}
on the numerical flux at the inner and outer edges of the simulation
box, where the second component of the subscripts,
  $j_{-}$ and $j_{+}$, denotes the $\phi$ locations at
$R_{-}$ ($i=1/2$) and $R_{+}$ ($i=n+1/2$), respectively. We note
that the relative position between $j_{-}$ and $j_{+}$ changes with
time according to the shearing boundary condition of
eq.(\ref{eq:shearcd}), which is a natural extension from to the
Cartesian shearing box setup \citep{hgb95}. 
We also note that the $\phi$ location that corresponds to $j_{\pm}$
does not generally coincide with the exact position of a fixed grid
cell because the shear evolves with time.   
Therefore, we need to interpolate the adjoining two cells along
the $\phi$ axis to derive $S^{m+\nicefrac{1}{2}}_{\nicefrac{1}{2},j_{-}}$ and
$S^{m+\nicefrac{1}{2}}_{n+\nicefrac{1}{2},j_{+}}$.

If eq.(\ref{eq:Ardsh}) is applied to $S_{\rm mass}$ (eq.\ref{eq:Amass}),
the total mass in the simulation box is conserved within round-off error,
as shown in eq. (\ref{eq:MassConsv}). The azimuthal magnetic flux at
shearing planes (eq.\ref{eq:phiphi}) and the vertical magnetic flux
at horizontal planes (eq.\ref{eq:phiz}) are conserved to round-off error
by applying eq.(\ref{eq:Ardsh}) to $S_{B_{\phi}}=cE_z$ (eq.\ref{eq:ABp}) and
$S_{B_z}=RcE_{\phi}$ (eq.\ref{eq:ABz}), respectively. We explain our specific
method for the magnetic fluxes in Appendix \ref{sec:shCT}. 

In addition to numerical fluxes, $F$, it is needed to apply the shearing
boundary condition to variables, $V$, located at the center of ghost cells,
in order to derive the numerical flux $F^{m+\nicefrac{1}{2}}$ at the
inner ($i=1/2$) and outer ($i=n+1/2$) boundaries of the simulation box.
$V_0$ ($V_{n+1}$) is also necessary to determine the slope of $V_1$
($V_n$) when the 2nd order spacial accuracy is required;
for higher-order accuracy than 2nd order, more than one ghost cell per
boundary needs to be prepared, {\it i.e.}, to achieve $(k+2)$-th order
accuracy, up to $S_{-k}(=S_{n-k})$ and $S_{n+k+1}(=S_{k+1})$ are
necessary to determine the slope of $S_{1}$ and $S_{n}$, respectively. 

We apply the shearing condition to cell centered values from the
innermost active ($i=1$) cells to the corresponding sheared outer
ghost ($i=n+1$) cells,  
\begin{equation}
  S^{m}_{n+1,j_{+g}} = S^{m}_{1,j_{-a}},
  \label{eq:shearvc1}
\end{equation}
and from the outermost active ($i=n$) cells to the corresponding inner
ghost ($i=0$) cells,
\begin{equation}
  S^{m}_{0,j_{-g}} = S^{m}_{n,j_{+a}}, 
  \label{eq:shearvc2}
\end{equation}
where we add ``a'' or ``g'' to the $\phi$ subscripts, $j_{\pm}$, to
explicitly show the active or ghost cell. 

As for $V=\rho$, $\rho v_R$, and $\rho v_z$, we can use the simple scaling
relations derived in eqs. (\ref{eq:smplsc1}) \& (\ref{eq:smplsc2}).
On the other hand, the other $V=\rho v_{\phi}$, $B_R$, $B_{\phi}$, $B_z$, and
$\frac{1}{2} \rho v^2 + \rho e + \frac{B^2}{8\pi}$ are not directly
connected to the shearing variables, $S$, via simple relations.
The most straightforward way is probably to iteratively derive
these five $V$ at the ghost cells from $S_{{\rm mom},\phi}$, $S_{\rm
  eng}$, $S_{B_\phi}$, and $S_{B_z}$ at the corresponding active cells
under the constraint of $\mbf{\nabla\cdot B}=0$. 

However, this procedure is not suited to a staggered mesh system in which
the three components of the magnetic field are located at different positions
from those of the other variables, because we need multiple interpolations,
which could reduce the numerical accuracy. 
Therefore, it is better to adopt a different strategy for the staggered mesh
system, as described below.

\subsection{Staggered Meshes}
\label{sec:shstg}

\begin{figure}
  \begin{center}
    \includegraphics[width=0.5\textwidth]{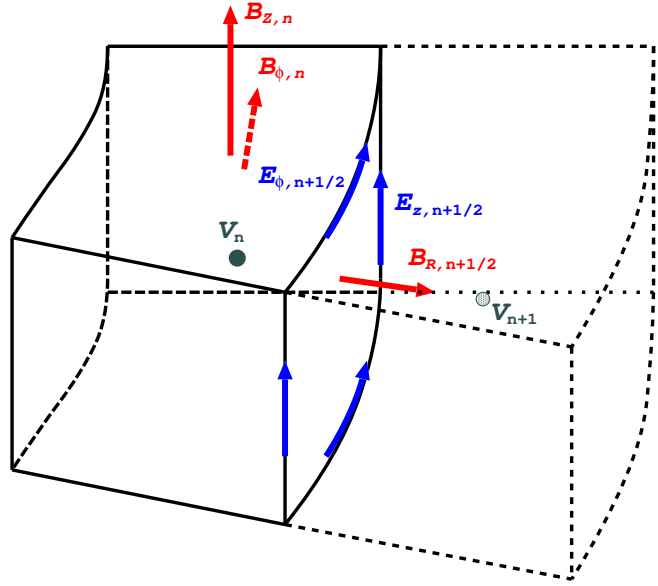}
  \end{center}
  \caption{Grid structure at the radial outer boundary, $R=R_{+}$.
    The outermost active cell labeled with subscript $n$ is drawn by
    solid lines and the ghost cell with $n+1$ is by dotted lines. Each 
    component of magnetic field is located at the corresponding
    surface of the cell, and the induced electric field,
    $\mbf{cE}=-\mbf{v\times B}$, is at the side. HD
    variables, represented by $V$, are at the center of the cell.
    \label{fig:cellstr}
  }
\end{figure}

The constraint transport (CT) method \citep{eh88} is a numerical
scheme to update magnetic fields under the constraint of
$\mbf{\nabla\cdot B}=0$ within the precision of round-off error.  
In the CT scheme, the three components of the magnetic field are placed on
the surfaces of each grid cell (Figure \ref{fig:cellstr}). On the
other hand, the HD variables are located at the center of
the cell. We apply the shearing periodic boundary presented in
Subsection \ref{sec:rbc} to these staggered meshes.    


\subsubsection{Primitive Variables}


Let us first explain how we apply the radial shearing boundary condition
to the primitive variables, $V=\rho$, $\mbf{v}$, $\mbf{B}$, and $e$, at
ghost cells and at time $t=m$ by eqs.(\ref{eq:shearvc1})
\& (\ref{eq:shearvc2}).
As for $\rho$, $v_{R}$, and $v_z$, we can use the simple scaling relations
of eqs. (\ref{eq:smplsc1}) \& (\ref{eq:smplsc2}):
\begin{equation}
  (\rho R)_{n+1,j_{+g}} = (\rho R)_{1,j_{-a}}; \; (\rho R)_{0,j_{-g}}
    = (\rho R)_{n,j_{+a}} , 
\end{equation}
\begin{equation}
   (v_R)_{n+1,j_{+g}} = (v_R)_{1,j_{-a}}; \; (v_R)_{0,j_{-g}} =
      (v_R)_{n,j_{+a}} , 
  \label{eq:shearvR}
\end{equation}
and
\begin{equation}
   (v_z)_{n+1,j_{+g}} = (v_z)_{1,j_{-a}}; \; (v_z)_{0,j_{-g}} =
    (v_z)_{n,j_{+a}} . 
  \label{eq:shearvz}
\end{equation}


We introduced the sum of Maxwell and Reynolds stresses for the angular
momentum shearing variable, $S_{{\rm mom},\phi}$ (eq.\ref{eq:Aangmom}) in
Subsection \ref{sec:rbc}. We utilize $S_{{\rm mom},\phi}$ to
determine $v_{\phi}$ and $B_{\phi}$ at the ghost cells. 
We here assume both HD and magnetic components have the same radial
scaling as that of eq. 
(\ref{eq:Aangmom}), namely
\begin{equation}
  \rho v_R \delta v_{\phi} \propto \Omega_{\rm eq}, 
  \label{eq:Ryshear}
\end{equation}
and
\begin{equation}
  B_R B_{\phi} \propto \Omega_{\rm eq}.
  \label{eq:Mxshear}
\end{equation}
Eqs. (\ref{eq:Ryshear}) \& (\ref{eq:Amass}) give
\begin{equation}
\Delta \Omega\equiv \frac{\delta v_{\phi}}{R} \propto \Omega_{\rm eq}(R),  
\end{equation}
and therefore,
\begin{equation}
  (\delta v_{\phi}/R\Omega_{\rm eq})_{n+1,j_{+g}} = (\delta
  v_{\phi}/R\Omega_{\rm eq})_{1,j_{-a}} 
  ; \;\;
  (\delta v_{\phi}/R\Omega_{\rm eq})_{0,j_{-g}} = (\delta
  v_{\phi}/R\Omega_{\rm eq})_{n,j_{+a}} .   
  \label{eq:shearvp}
\end{equation}

\begin{figure}
  \begin{center}
    \includegraphics[width=0.7\textwidth]{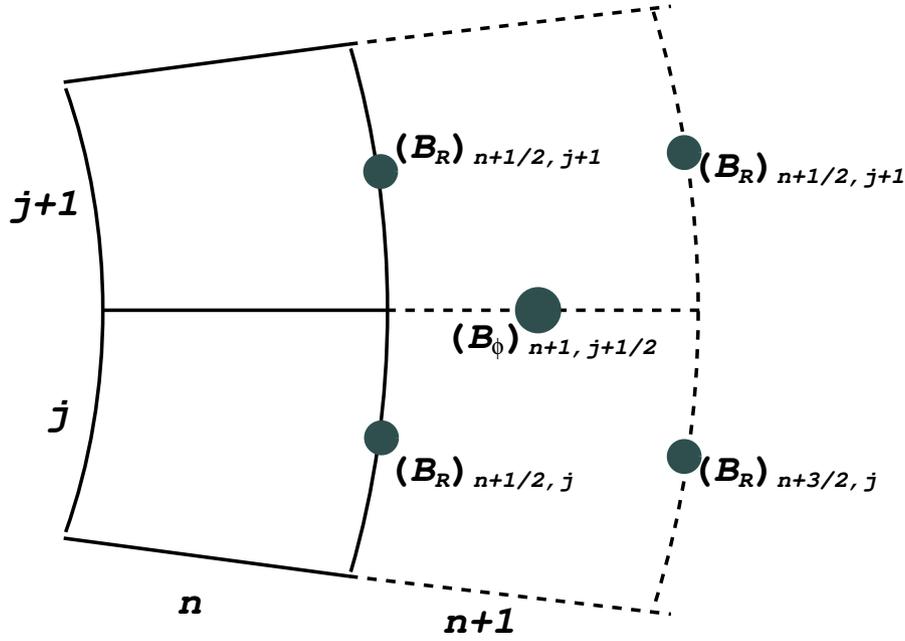}
  \end{center}
  \caption{Locations of $B_R$ and $B_{\phi}$ (gray dots) near the outer
    boundary on a horizontal plane. For simplicity, we
      write $j$ here, though it should be $j_{+g}$ in a strict sense,
      following eq.(\ref{eq:shearvc1}).
    \label{fig:BRBplocation}
  }
\end{figure}

When we apply eq.(\ref{eq:Mxshear}) to the staggered meshes, we need
to interpolate because the locations of $B_R$ and $B_{\phi}$ are
different, as shown in Figure \ref{fig:BRBplocation}. The shearing condition
of eq.(\ref{eq:Mxshear}) is applied at the location of $B_{\phi}$ as
follows: 
\begin{equation}
  \left(B_R B_{\phi}/\Omega_{\rm
    eq}\right)_{n+1,(j_{+g})+\nicefrac{1}{2}} 
  = \left(B_R B_{\phi}/\Omega_{\rm
    eq}\right)_{1,(j_{-a})+\nicefrac{1}{2}}; \;\; 
  \left(B_R B_{\phi}/\Omega_{\rm
    eq}\right)_{0,(j_{-g})+\nicefrac{1}{2}} 
  = \left(B_R B_{\phi}/\Omega_{\rm
    eq}\right)_{n,(j_{+a})+\nicefrac{1}{2}}. 
  \label{eq:BpBrsh}
\end{equation}

We also need an interpolation for $B_R$ in eq.(\ref{eq:BpBrsh}), because
$B_R$ is located at different positions from that of $B_{\phi}$
(Figure \ref{fig:BRBplocation}). We take the simple average of the four
neighboring locations to calculate $B_{R}$ at $i=1$ \& $n$:
\begin{eqnarray}
  & &\left(B_{R}\right)_{1,(j_{-a})+\nicefrac{1}{2}} = \frac{1}{4}
  \left[(B_R)_{\nicefrac{1}{2},(j_{-a})} + (B_R)_{\nicefrac{1}{2},(j_{-a})+1}
    + (B_R)_{\nicefrac{3}{2},(j_{-a})} +
    (B_R)_{\nicefrac{3}{2},(j_{-a})+1}\right] \nonumber \\ 
  & &\left(B_{R}\right)_{n,(j_{+a})+\nicefrac{1}{2}} =
  \frac{1}{4}\left[(B_R)_{n-\nicefrac{1}{2},(j_{+a})} 
    + (B_R)_{n-\nicefrac{1}{2},(j_{+a})+1} + (B_R)_{n+\nicefrac{1}{2},(j_{+a})}
    + (B_R)_{n+\nicefrac{1}{2},(j_{+a})+1}\right].
  \label{eq:Br4av1}
\end{eqnarray}

On the other hand, we have to carefully deal with $B_R$ at the ghost cells
to avoid numerical cancellation, which causes spurious behavior of $B_{\phi}$.
First, we take the simple average of the four neighboring locations
in the same manner to eq.(\ref{eq:Br4av1}):
\begin{eqnarray}
  & &\left(B_{R,{\rm av}}\right)_{n+1,(j_{+g})+\nicefrac{1}{2}} =
  \frac{1}{4}\left[(B_R)_{n+\nicefrac{1}{2},(j_{+g})} 
    + (B_R)_{n+\nicefrac{1}{2},(j_{+g})+1} + (B_R)_{n+\nicefrac{3}{2},(j_{+g})} +
    (B_R)_{n+\nicefrac{3}{2},(j_{+g})+1}\right] \nonumber \\ 
  & &\left(B_{R,{\rm av}}\right)_{0,(j_{-g})+\nicefrac{1}{2}} =
  \frac{1}{4}\left[(B_R)_{-\nicefrac{1}{2},(j_{-g})} +
    (B_R)_{-\nicefrac{1}{2},(j_{-g})+1} + 
    (B_R)_{\nicefrac{1}{2},(j_{-g})} +
    (B_R)_{\nicefrac{1}{2},(j_{-g})+1}\right],   
  \label{eq:Br4av2}
\end{eqnarray}
where $(B_R)_{n+\nicefrac{3}{2},(j_{+g})}$,
$(B_R)_{n+\nicefrac{3}{2},(j_{+g})+1}$,
$(B_R)_{-\nicefrac{1}{2},(j_{-g})}$, and
$(B_R)_{-\nicefrac{1}{2},(j_{-g})+1}$ are still unknown. We here use
\begin{eqnarray}
  & &(B_R R)_{n+\nicefrac{3}{2},(j_{+g})} = (B_R
  R)_{\nicefrac{3}{2},(j_{-a})};\; (B_R R)_{n+\nicefrac{3}{2},(j_{+g})+1} = (B_R
  R)_{\nicefrac{3}{2},(j_{-a})+1}\nonumber \\
  & &(B_R R)_{-\nicefrac{1}{2},(j_{-g})}
  = (B_R R)_{n-\nicefrac{1}{2},(j_{+a})};\; (B_R R)_{-\nicefrac{1}{2},(j_{-g})+1}
  = (B_R R)_{n-\nicefrac{1}{2},(j_{+a})+1}  
  \label{eq:BRsc}
\end{eqnarray}
which are expected from the radial differential term of
$\mbf{\nabla\cdot B}=0$.
If the signs of $B_{R}$ on the right-hand side of eq.(\ref{eq:Br4av2}) are
different, numerical cancellation occasionally occurs to give
$(B_{R,{\rm av}})_{n+1,(j_{+g})+\nicefrac{1}{2}}$ or $(B_{R,{\rm
    av}})_{0,(j_{-g})+\nicefrac{1}{2}}$ $\approx 0$, even though all
four $B_{R}$ on the right-hand side have finite values. 
If this is the case, applying $(B_{R})_{n+1,(j_{+g})+\nicefrac{1}{2}}=(B_{R,{\rm
    av}})_{n+1,(j_{+g})+\nicefrac{1}{2}}$ or
$(B_{R})_{0,(j_{-g})+\nicefrac{1}{2}}=(B_{R,{\rm
    av}})_{0,(j_{-g})+\nicefrac{1}{2}}$ to
eq.(\ref{eq:BpBrsh}) would give a spuriously huge absolute value of
$(B_{\phi})_{n+1,(j_{+g})+\nicefrac{1}{2}}$ or
$(B_{\phi})_{0,(j_{-g})+\nicefrac{1}{2}}$.

In order to avoid this unphysical behavior, we set a floor, $B_{R,{\rm min}}$,
on the interpolated $B_R$: 
\begin{eqnarray}
  & &(B_R)_{n+1,(j_{+g})+\nicefrac{1}{2}} = \mathrm{sgn} \left[
    (B_{R,{\rm  av}})_{n+1,(j_{+g})+\nicefrac{1}{2}} \right]\times
  \max\left[|(B_{R,{\rm av}})_{n+1,(j_{+g})+\nicefrac{1}{2}}|, (B_{R,{\rm
        min}})_{n+1,(j_{+g})+\nicefrac{1}{2}}\right] \nonumber \\ 
  & &(B_R)_{0,(j_{-g})+\nicefrac{1}{2}} = \mathrm{sgn} \left[
    (B_{R,{\rm av}})_{0,(j_{-g})+\nicefrac{1}{2}}\right]\times
  \max\left[|(B_{R,{\rm av}})_{0,(j_{-g})+\nicefrac{1}{2}}|, (B_{R,{\rm
        min}})_{0,(j_{-g})+\nicefrac{1}{2}}\right]. 
  \label{eq:BRchoice}
\end{eqnarray}
For $B_{R,{\rm min}}$ we take the minimum absolute value of the
four neighboring $B_{R}$ multiplied by a factor, $f_{R,{\rm min}}$, of
order of unity: 
\begin{eqnarray}
  & &(B_{R,{\rm min}})_{n+1,(j_{+g})+\nicefrac{1}{2}}= f_{R,{\rm min}}\times
  \min\left(|(B_R)_{n+\nicefrac{1}{2},(j_{+g})}|,|(B_R)_{n+\nicefrac{1}{2},(j_{+g})+1}| 
  ,|(B_R)_{n+\nicefrac{3}{2},(j_{+g})}|,|(B_R)_{n+\nicefrac{3}{2},(j_{+g})+1}| 
  \right) \nonumber \\
  & &(B_{R,{\rm min}})_{0,(j_{-g})+\nicefrac{1}{2}}= f_{R,{\rm min}}\times \min\left(
  |(B_R)_{-\nicefrac{1}{2},(j_{-g})}|,|(B_R)_{-\nicefrac{1}{2},(j_{-g})+1}|,|(B_R)_{\nicefrac{1}{2},(j_{-g})}|,|(B_R)_{\nicefrac{1}{2},(j_{-g})+1}|\right)
\end{eqnarray}
When $B_{R,{\rm min}}$ is selected in eq.(\ref{eq:BRchoice}) at a
ghost cell, the derived $B_{\phi}$ depends on the choice of $f_{R,{\rm
    min}}$. 
Accordingly, $f_{R,{\rm min}}$ controls the magnetic pressure
across the simulation boundary, $-\partial_R (B_{\phi}^2/8\pi)$, at
the ghost cell. As a result, the accretion velocity, $v_R$, also
depends on $f_{R,{\rm min}}$. We carefully determine $f_{R,{\rm min}}$
to give the global radial balance of the angular momentum flux between
mass accretion the MHD turbulence that was discussed in Sub-subsection
\ref{sec:angmomaccr}. We adopt $f_{R,{\rm min}}=1/\sqrt{2}$ in the
simulation we presented in this paper. 


We do not directly use shearing variables for $B_z$, but take a simple
assumption that the initial radial profile is preserved. Then,
$B_z$ at the ghost cells are determined by 
\begin{equation}
  (B_z R^{q_B})_{n+1,j_{+g}} = (B_z R^{q_B})_{1,j_{-a}}; \;\; (B_z R^{q_B})_{0,j_{-g}} = (B_z R^{q_B})_{n,j_{+a}}.
  \label{eq:Bzsc}
\end{equation}
In this paper, we adopted $q_{B}=1$, which gives the consistent
radial scalings of $v_R$ (eq.\ref{eq:shearvR}), $v_z$ (eq.\ref{eq:shearvz}),
$B_R$ (eq.\ref{eq:BRsc}), and $B_z$
(eq.\ref{eq:Bzsc}) with $S_{B_z}=RcE_{\phi}$ (eq.\ref{eq:ABz}). 

Although we do not solve an energy equation, for completeness we describe
how $e$ is determined at the ghost cells. From $S_{\rm eng}$ (eq.\ref{eq:Aeng})
and $S_{\rm mass}$ (eq.\ref{eq:Amass}), we obtain
\begin{equation}
  \left(\frac{v^2}{2} + (\gamma -1)e\right)_{n+1,j_{+g}} = \left(\frac{v^2}{2}
  + (\gamma -1)e\right)_{1,j_{-a}}; \;\;
  \left(\frac{v^2}{2} + (\gamma -1)e\right)_{0,j_{-g}} = \left(\frac{v^2}{2}
  + (\gamma -1)e\right)_{n,j_{+a}} .   
  \label{eq:sheare}
\end{equation}
All the three components of $\mbf{v}$ are already derived by eqs.
(\ref{eq:shearvR}), (\ref{eq:shearvz}), \& (\ref{eq:shearvp}), and therefore,
from eq.(\ref{eq:sheare}) we can determine $e$ at the ghost cells. 

\subsubsection{Numerical Fluxes}
\label{sec:shCT}
By using the variables, $V^{m}$, at the ghost cells, we can derive
the numerical flux, $F^{m+\nicefrac{1}{2}}=S^{m+\nicefrac{1}{2}}$, at
the simulation boundaries ($i=1/2$ and $n+1/2$) in
eq.(\ref{eq:fdeq}). However, the 
calculated $S^{m+1/2}$ does not guarantee that eq.(\ref{eq:Ardsh})
will be within the precision of round-off error because of the azimuthal
interpolation at the shearing boundary. In order to conserve the
invariant quantities introduced in Subsection \ref{sec:cq} within a
round-off error, it is necessary to apply corrections to the derived
$S^{m+\nicefrac{1}{2}}$.  

When we apply the shearing periodic condition to the magnetic field, we use
$S_{B_{\phi}} = RcE_{\phi}$ (eq.\ref{eq:ABp}) and $S_{B_z}=cE_z$ (eq.\ref{eq:ABz}),
which are the induced electric fields located at the exact radial
boundaries of the simulation box (Figure
\ref{fig:cellstr}). $E_{\phi}$ and $E_z$ 
at the radial boundaries are related to the conservation of magnetic flux, 
as we discussed in Subsection \ref{sec:cq}.

In order to conserve the vertical magnetic flux through $z$ planes
(eq.\ref{eq:phiz}) to round-off error, the line integration of
$E_{\phi}$ along the $\phi$ axis at $R_{-}$ and at $R_{+}$ must be equal: 
\begin{equation}
  \int_{\phi_{-}}^{\phi_{+}}d\phi (R E_{\phi})_{-} =
  \int_{\phi_{-}}^{\phi_{+}}d\phi (R E_{\phi})_{+},
  \label{eq:Epbd}
\end{equation}
where subscript `$-$' corresponds to $i=1/2$ and `$+$' to $i=n+1/2$.
$E_{\phi}$ at the radial boundaries are evaluated from the boundary
cell ($i=1$ or $n$) and the ghost cell ($i=0$ or $n+1$), and they
do not usually satisfy the above conservation relation, as previously
discussed.
We take the average of the original value of $E_{\phi}$ at $R_{\pm}$
and $E_{\phi}$ at $R_{\mp}$ at the corresponding sheared location: 
\begin{eqnarray}
  & &(R E_{\phi}^{\rm cr}(\phi))_{-} = \frac{1}{2} \left[ (R
    E_{\phi}(\phi))_{-} + (RE_{\phi}(\phi-\Delta\Omega
    t))_{+}\right] \nonumber \\
  & &(R E_{\phi}^{\rm cr}(\phi))_{+} = \frac{1}{2}\left[ (R
    E_{\phi}(\phi+\Delta\Omega t))_{-} + (RE_{\phi}(\phi))_{+}\right], 
  \label{eq:Epave}
\end{eqnarray}
or in the discretized forms, 
\begin{eqnarray}
  & &(R E_{\phi}^{\rm cr})_{\nicefrac{1}{2},j_{-}} =
  \frac{1}{2}\left[(R E_{\phi})_{\nicefrac{1}{2},j_{-}} + (R
    E_{\phi})_{n+\nicefrac{1}{2},j_{+}} \right] \nonumber \\ 
  & & (R E_{\phi}^{\rm cr})_{n+\nicefrac{1}{2},j_{+}} =
  \frac{1}{2}\left[(R E_{\phi})_{n+\nicefrac{1}{2},j_{+}} 
    + (R E_{\phi})_{\nicefrac{1}{2},j_{-}} \right] ,
  \label{eq:Epave2}
\end{eqnarray}
The position of $j_{-}$ and $j_{+}$ does not usually match a grid
cell, and therefore, the azimuthal interpolation is necessary to derive
$(E_{\phi})_{\nicefrac{1}{2},j_{-}}$ and
$(E_{\phi})_{n+\nicefrac{1}{2},j_{+}}$.  
We use a simple linear interpolation, which is sufficient to satisfy
the conservation relation of eq.(\ref{eq:Epbd}). 

When updating the magnetic fields, we use $E_{\phi}^{\rm cr}(\phi)$,
instead of $E_{\phi}(\phi)$, at the $R_{\pm}$ boundaries of $i=1/2$
and $n+1/2$. This correction ensures that the vertical magnetic flux
is conserved (eq.\ref{eq:phiz}) within the round-off error according
to eq.(\ref{eq:Epbd}).   

Similar to the relation between $E_{\phi}$ and $B_z$, $E_z(=(v_R B_{\phi}
- v_{\phi}B_R)/c)$ at the $R_{\pm}$ boundaries regulates the conservation
of azimuthal magnetic flux (eq.\ref{eq:phiphi}). More specifically,
\begin{equation}
  \int_{z_{-}}^{z_{+}}dz (E_z)_{+}=\int_{z_{-}}^{z_{+}}dz (E_z)_{-} 
  \label{eq:Ezz}
\end{equation}
conserves the azimuthal magnetic flux through shearing planes
(eq.\ref{eq:phiphi}), where 
the $z$ integral is taken at the radial boundaries of each shearing plane.

We slightly modify the correction method for $E_{\phi}$
(eqs.\ref{eq:Epave} \& \ref{eq:Epave2}) in order to apply it to
$E_{z}$ because $v_{\phi}$ in $E_z$ contains the mean rotational 
velocity that has opposite signs at $R_{+}$ and
$R_{-}$. $v_{\phi}B_R$ in $E_z$ at the two corresponding
sheared locations of $R_{-}$ and $R_{+}$ could have very different
values. In this case, if we take the local average of the two
corresponding sheared locations, as done for $E_{\phi}$ 
(eq.\ref{eq:Epave}), it may cause spurious numerical errors.

Instead of taking the local average, we use the integrated average of
$E_z$ over the $\phi z$ planes at $R_{\pm}$ to derive a correction,
\begin{eqnarray}
  & &(E_z^{\rm cr})_{-} = (E_z)_{-} + \frac{1}{2}\left[\overline{(E_{z})}_{+}
    - \overline{(E_{z})}_{-}\right] \nonumber \\
  & &(E_z^{\rm cr})_{+} = (E_z)_{+} + \frac{1}{2}\left[\overline{(E_{z})}_{-}
    - \overline{(E_{z})}_{+}\right],
  \label{eq:correcEz}
\end{eqnarray}
where in the discretized form,
$(E_z)_{-}=(E_z)_{\nicefrac{1}{2},j_{-}}$ and
$(E_z)_{+}=(E_z)_{n+\nicefrac{1}{2},j_{+}}$, and 
$\overline{(E_{z})}_{\pm}$ is the integrated average, 
\begin{equation}
  \overline{(E_{z})}_{\pm} = \frac{\int_{\phi_{-}}^{\phi_{+}}d\phi
    \int_{z_{-}}^{z_{+}}dz (E_z)_{\pm}}{\int_{\phi_{-}}^{\phi_{+}}d\phi
    \int_{z_{-}}^{z_{+}}dz}.
  \label{eq:Ezintave}
\end{equation}
By taking the global average, $\overline{(E_{z})}_{\pm}$, random
differences between the two $E_z$'s at the corresponding sheared
locations of $R_{\pm}$ can be canceled out. Therefore, we can reduce
spurious errors of the correction when taking the local average by
using eq.(\ref{eq:correcEz}). 

One may notice that in eq.(\ref{eq:Ezintave}) only the $z$ integration
along both radial boundaries of a shearing plane is
sufficient to satisfy the conservation of $\Phi_{\phi}$ from
eq.(\ref{eq:Ezz}). However, the locations of the radial boundaries do
not generally match grid cells, and therefore, the $\phi$
interpolation is required to match the time-evolving shearing planes at
each time step. It is simpler to take the $\phi$ average without $\phi$
interpolation. Moreover, random errors can further be
canceled out by the $\phi$ integration, in addition to the $z$
integration. Therefore, we take both $\phi$ and $z$ integration to
derive $ \overline{(E_{z})}_{\pm}$.

It is also difficult to check the conservation of $\Phi_{\phi}$ at
shearing planes by the same reason explained above. When we
numerically test the conservation of $\Phi_{\phi}$, we also check the
conservation of $\sum_{\phi}\Phi_{\phi}$. 

In our simulations, we implement corrections of the numerical
fluxes only in the CT scheme of $E_{\phi}$ and $E_z$.
If one likes to apply eq.(\ref{eq:Ardsh}) to $S_{\rm mass}$ for mass
conservation within the round-off error, the same procedure for
$E_{\phi}$ (eqs.\ref{eq:Epave} \& \ref{eq:Epave2}) can be adopted.

\section{Epicyclic Oscillation}
\label{sec:epiosc}
We derive eqs.(\ref{eq:epiR}) \& (\ref{eq:epiphi}) from the
cylindrical shearing box formulation. We neglect the magnetic terms
below. The radial component of the momentum flux averaged over the
$\phi$ and $z$ directions is  
\begin{eqnarray}
\frac{\partial}{\partial t}\langle\rho v_R\rangle +
\frac{1}{R}\frac{\partial}{\partial R}\langle\rho v_R R^2\rangle
&\approx& -\frac{\partial \langle p\rangle}{\partial R} + \frac{\langle\rho
  v_{\phi}^2\rangle}{R} -\langle\rho\rangle\frac{GM_{\star}}{R^2} +
\langle\rho\rangle R \Omega_{\rm eq,0}^2 + 2\Omega_{\rm eq,0}\langle\rho
v_{\phi}\rangle \nonumber \\
&=& -\frac{\partial \langle \delta p\rangle}{\partial R}
+ 2\left(\Omega_{\rm eq,0} + \frac{v_{\phi,{\rm eq}}}{R}\right)
\langle\rho \delta v_{\phi}\rangle + \frac{\langle\rho\delta
  v_{\phi}^2\rangle}{R} \nonumber \\ 
&\approx& 2\Omega_{\rm eq,0}\langle\rho \delta v_{\phi}\rangle,
\label{eq:epiRapp}
\end{eqnarray}
where $\delta p = p - p_{\rm eq}$ and we refer to eq.(\ref{eq:exterm})
when deriving the second equality. We leave the dominant term of the
right-hand side of the second equality to obtain the final
expression. The volume integral of eq.(\ref{eq:epiRapp}) gives
eq.(\ref{eq:epiR}). 

The azimuthal component of the $\phi$ and $z$ averaged momentum flux
is 
\begin{equation}
  \frac{\partial}{\partial t}\langle\rho v_{\phi}R\rangle +
  \frac{1}{R}\frac{\partial}{\partial R}\langle\rho v_{\phi}v_R
  R^2\rangle \approx -2\Omega_{\rm eq,0}\langle\rho v_R R\rangle. 
  \label{eq:epiphiapp}
\end{equation}
We take the volume integral of this equation. The second term on the
left-hand side is integrated as 
\begin{equation}
  \int_{R_{-}}^{R_{+}}\frac{1}{R}\frac{\partial}{\partial R}\langle\rho
    v_{\phi}v_R R^2\rangle R dR = \langle \rho v_R R\rangle \left[R
      v_{\phi}\right]_{R{-}}^{R_{+}}, 
    \label{eq:epiphi2vol}
\end{equation}
where we factored out the shearing variable of $S_{\rm mass}=\rho v_R
R$ from the integration. We can expand
$R_{\pm}^2\approx R_{0}^2\left(1+\frac{2(R_{\pm}-R_0)}{R_0}\right)$ and
$\Omega_{\rm eq,\pm}\approx \Omega_{\rm eq,0}\left(1 \mp
\frac{3}{2}\frac{R_{\pm}-R_0}{R_0}\right)$ for $(R_{+}-R_{-})\ll R_0$
and $H_0\ll R_0$. Then, $\left[Rv_{\phi}\right]_{R{-}}^{R_{+}}$ can be
written as
\begin{eqnarray}
  (Rv_{\phi})_{+} - (Rv_{\phi})_{-} &=& R_{+}^2(\Omega_{\rm eq,+}
  -\Omega_{\rm eq,0}) - R_{-}^2(\Omega_{\rm eq,-}-\Omega_{\rm eq,0})
  + (R\delta v_{\phi})_{+} - (R\delta v_{\phi})_{-} \nonumber \\
    &\approx& -\frac{3}{2}R_{0}\Omega_{\rm eq,0}(R_{+} - R_{-})
      -\frac{1}{2}R_{0}\Omega_{\rm eq,0}\Delta (R_{+} - R_{-})
      \approx - \Omega_{\rm eq,0}\left(\frac{3}{2} +
      \frac{\Delta}{2}\right)\int_{R_{-}}^{R_{+}}RdR,
      \label{eq:epiphi3vol}
\end{eqnarray}
where $\Delta \equiv \delta \Omega_{+} / \Omega_{\rm eq,+} = \delta
\Omega_{-} / \Omega_{\rm eq,-}$ and we used $R_0 \approx 
\frac{1}{2}(R_{-}+R_{+})$ to derive the final expression.  
Applying eqs.(\ref{eq:epiphi2vol}) \& (\ref{eq:epiphi3vol}) to the
volume integral of eq.(\ref{eq:epiphiapp}), we have
\begin{equation}
  \frac{\partial}{\partial t}\int_{R_{-}}^{R_{+}}RdR\langle\rho
  v_{\phi}R\rangle   - \Omega_{\rm eq,0}\left(\frac{3}{2} +
      \frac{\Delta}{2}\right)\int_{R_{-}}^{R_{+}}RdR\langle\rho v_R
      R\rangle = -2\Omega_{\rm eq,0}\int_{R_{-}}^{R_{+}}RdR\langle\rho
      v_R R\rangle,  
\end{equation}
which is further transformed into 
\begin{equation}
  \frac{\partial}{\partial t}\left[\rho v_{\phi}R\right]_{V} =
  -\frac{1}{2}\left(1-\Delta\right)\Omega_{\rm eq,0}\left[\rho v_R
    R\right]_{V}. 
\end{equation}
We can usually assume $\Delta \ll 1$ and 
$\frac{\partial}{\partial t}[\rho v_{\phi}R]_V
= \frac{\partial}{\partial t}[(\rho v_{\phi,{\rm eq}}
+ \rho \delta v_{\phi})R]_V
\approx \frac{\partial}{\partial t}[\rho \delta v_{\phi}R]_V$, which
give eq.(\ref{eq:epiphi}).

\end{appendix}

\twocolumn
\bibliography{paper_cylsh2018}

\begin{thebibliography}{95}
\expandafter\ifx\csname natexlab\endcsname\relax\def\natexlab#1{#1}\fi

\bibitem[{{Adachi} {et~al.}(1976){Adachi}, {Hayashi}, \& {Nakazawa}}]{ada76}
{Adachi}, I., {Hayashi}, C., \& {Nakazawa}, K. 1976, Progress of Theoretical
  Physics, 56, 1756

\bibitem[{{Armitage}(1998)}]{arm98}
{Armitage}, P.~J. 1998, \apjl, 501, L189

\bibitem[{{Bai}(2013)}]{bai13}
{Bai}, X.-N. 2013, \apj, 772, 96

\bibitem[{{Bai} \& {Stone}(2013{\natexlab{a}})}]{bs13a}
{Bai}, X.-N., \& {Stone}, J.~M. 2013{\natexlab{a}}, \apj, 767, 30

\bibitem[{{Bai} \& {Stone}(2013{\natexlab{b}})}]{bs13b}
---. 2013{\natexlab{b}}, \apj, 769, 76

\bibitem[{{Balbus} \& {Hawley}(1991)}]{bh91}
{Balbus}, S.~A., \& {Hawley}, J.~F. 1991, \apj, 376, 214

\bibitem[{{Balbus} \& {Hawley}(1998)}]{bh98}
---. 1998, Reviews of Modern Physics, 70, 1

\bibitem[{{Baruteau} {et~al.}(2014){Baruteau}, {Crida}, {Paardekooper},
  {Masset}, {Guilet}, {Bitsch}, {Nelson}, {Kley}, \& {Papaloizou}}]{bar14}
{Baruteau}, C., {Crida}, A., {Paardekooper}, S.-J., {et~al.} 2014, Protostars
  and Planets VI, 667

\bibitem[{{Ben{\'{\i}}tez-Llambay} \& {Masset}(2016)}]{fargo16}
{Ben{\'{\i}}tez-Llambay}, P., \& {Masset}, F.~S. 2016, \apjs, 223, 11

\bibitem[{{Blackman} \& {Nauman}(2015)}]{bn15}
{Blackman}, E.~G., \& {Nauman}, F. 2015, Journal of Plasma Physics, 81,
  395810505

\bibitem[{{Blandford} \& {Payne}(1982)}]{bp82}
{Blandford}, R.~D., \& {Payne}, D.~G. 1982, \mnras, 199, 883

\bibitem[{{Brandenburg} {et~al.}(1995){Brandenburg}, {Nordlund}, {Stein}, \&
  {Torkelsson}}]{bra95}
{Brandenburg}, A., {Nordlund}, A., {Stein}, R.~F., \& {Torkelsson}, U. 1995,
  \apj, 446, 741

\bibitem[{{Brandenburg} {et~al.}(1996){Brandenburg}, {Nordlund}, {Stein}, \&
  {Torkelsson}}]{bra96}
---. 1996, \apjl, 458, L45

\bibitem[{{Carballido} {et~al.}(2006){Carballido}, {Fromang}, \&
  {Papaloizou}}]{car06}
{Carballido}, A., {Fromang}, S., \& {Papaloizou}, J. 2006, \mnras, 373, 1633

\bibitem[{{Chandrasekhar}(1961)}]{cha61}
{Chandrasekhar}, S. 1961, {Hydrodynamic and hydromagnetic stability} (Oxford:
  Clarendon)

\bibitem[{{Clarke}(1996)}]{cl96}
{Clarke}, D.~A. 1996, \apj, 457, 291

\bibitem[{{Crida} \& {Morbidelli}(2007)}]{cm07}
{Crida}, A., \& {Morbidelli}, A. 2007, \mnras, 377, 1324

\bibitem[{{Davis} {et~al.}(2010){Davis}, {Stone}, \& {Pessah}}]{dav10}
{Davis}, S.~W., {Stone}, J.~M., \& {Pessah}, M.~E. 2010, \apj, 713, 52

\bibitem[{{Evans} \& {Hawley}(1988)}]{eh88}
{Evans}, C.~R., \& {Hawley}, J.~F. 1988, \apj, 332, 659

\bibitem[{{Ferreira} {et~al.}(2006){Ferreira}, {Dougados}, \& {Cabrit}}]{fer06}
{Ferreira}, J., {Dougados}, C., \& {Cabrit}, S. 2006, \aap, 453, 785

\bibitem[{{Flock} {et~al.}(2011){Flock}, {Dzyurkevich}, {Klahr}, {Turner}, \&
  {Henning}}]{flo11}
{Flock}, M., {Dzyurkevich}, N., {Klahr}, H., {Turner}, N.~J., \& {Henning}, T.
  2011, \apj, 735, 122

\bibitem[{{Fromang} {et~al.}(2013){Fromang}, {Latter}, {Lesur}, \&
  {Ogilvie}}]{fro13}
{Fromang}, S., {Latter}, H., {Lesur}, G., \& {Ogilvie}, G.~I. 2013, \aap, 552,
  A71

\bibitem[{{Fromang} \& {Papaloizou}(2007)}]{fp07}
{Fromang}, S., \& {Papaloizou}, J. 2007, \aap, 476, 1113

\bibitem[{{Gammie}(2001)}]{gam01}
{Gammie}, C.~F. 2001, \apj, 553, 174

\bibitem[{{Gressel} {et~al.}(2012){Gressel}, {Nelson}, \& {Turner}}]{gre12}
{Gressel}, O., {Nelson}, R.~P., \& {Turner}, N.~J. 2012, \mnras, 422, 1140

\bibitem[{{Guilet} \& {Ogilvie}(2012)}]{go12}
{Guilet}, J., \& {Ogilvie}, G.~I. 2012, \mnras, 424, 2097

\bibitem[{{Hawley}(2000)}]{haw00}
{Hawley}, J.~F. 2000, \apj, 528, 462

\bibitem[{{Hawley} {et~al.}(1995){Hawley}, {Gammie}, \& {Balbus}}]{hgb95}
{Hawley}, J.~F., {Gammie}, C.~F., \& {Balbus}, S.~A. 1995, \apj, 440, 742

\bibitem[{{Hayashi}(1981)}]{hay81}
{Hayashi}, C. 1981, Progress of Theoretical Physics Supplement, 70, 35

\bibitem[{{Hirose} {et~al.}(2006){Hirose}, {Krolik}, \& {Stone}}]{hir06}
{Hirose}, S., {Krolik}, J.~H., \& {Stone}, J.~M. 2006, \apj, 640, 901

\bibitem[{{Hirose} \& {Shi}(2019)}]{hs19}
{Hirose}, S., \& {Shi}, J.-M. 2019, \mnras, 485, 266

\bibitem[{{Hoshino}(2015)}]{hos15}
{Hoshino}, M. 2015, Physical Review Letters, 114, 061101

\bibitem[{{Io} \& {Suzuki}(2014)}]{is14}
{Io}, Y., \& {Suzuki}, T.~K. 2014, \apj, 780, 46

\bibitem[{{Jiang} {et~al.}(2013){Jiang}, {Stone}, \& {Davis}}]{jia13}
{Jiang}, Y.-F., {Stone}, J.~M., \& {Davis}, S.~W. 2013, \apj, 778, 65

\bibitem[{{Johansen} {et~al.}(2006){Johansen}, {Klahr}, \& {Henning}}]{joh06}
{Johansen}, A., {Klahr}, H., \& {Henning}, T. 2006, \apj, 636, 1121

\bibitem[{{Johansen} {et~al.}(2009){Johansen}, {Youdin}, \& {Klahr}}]{joh09}
{Johansen}, A., {Youdin}, A., \& {Klahr}, H. 2009, \apj, 697, 1269

\bibitem[{{Kanagawa} {et~al.}(2018){Kanagawa}, {Tanaka}, \&
  {Szuszkiewicz}}]{kan18}
{Kanagawa}, K.~D., {Tanaka}, H., \& {Szuszkiewicz}, E. 2018, \apj, 861, 140

\bibitem[{{Kimura} {et~al.}(2016){Kimura}, {Toma}, {Suzuki}, \&
  {Inutsuka}}]{kim16}
{Kimura}, S.~S., {Toma}, K., {Suzuki}, T.~K., \& {Inutsuka}, S.-i. 2016, \apj,
  822, 88

\bibitem[{{Kimura} {et~al.}(2019){Kimura}, {Tomida}, \& {Murase}}]{kim19}
{Kimura}, S.~S., {Tomida}, K., \& {Murase}, K. 2019, \mnras, 485, 163

\bibitem[{{Klahr} \& {Bodenheimer}(2003)}]{kla03}
{Klahr}, H.~H., \& {Bodenheimer}, P. 2003, \apj, 582, 869

\bibitem[{{Kobayashi} {et~al.}(2016){Kobayashi}, {Tanaka}, \&
  {Okuzumi}}]{kob16}
{Kobayashi}, H., {Tanaka}, H., \& {Okuzumi}, S. 2016, \apj, 817, 105

\bibitem[{{Kunz} \& {Lesur}(2013)}]{kun13}
{Kunz}, M.~W., \& {Lesur}, G. 2013, \mnras, 434, 2295

\bibitem[{{Kunz} {et~al.}(2016){Kunz}, {Stone}, \& {Quataert}}]{kun16}
{Kunz}, M.~W., {Stone}, J.~M., \& {Quataert}, E. 2016, Physical Review Letters,
  117, 235101

\bibitem[{{Latter} {et~al.}(2015){Latter}, {Fromang}, \& {Faure}}]{lat15}
{Latter}, H.~N., {Fromang}, S., \& {Faure}, J. 2015, \mnras, 453, 3257

\bibitem[{{Lazarian} \& {Vishniac}(1999)}]{lv99}
{Lazarian}, A., \& {Vishniac}, E.~T. 1999, \apj, 517, 700

\bibitem[{{Lesur} {et~al.}(2013){Lesur}, {Ferreira}, \& {Ogilvie}}]{les13}
{Lesur}, G., {Ferreira}, J., \& {Ogilvie}, G.~I. 2013, \aap, 550, A61

\bibitem[{{Li} {et~al.}(2011){Li}, {Krasnopolsky}, \& {Shang}}]{li11}
{Li}, Z.-Y., {Krasnopolsky}, R., \& {Shang}, H. 2011, \apj, 738, 180

\bibitem[{{Lubow} {et~al.}(1994){Lubow}, {Papaloizou}, \& {Pringle}}]{lub94}
{Lubow}, S.~H., {Papaloizou}, J.~C.~B., \& {Pringle}, J.~E. 1994, \mnras, 267,
  235

\bibitem[{{Lynden-Bell} \& {Pringle}(1974)}]{lp74}
{Lynden-Bell}, D., \& {Pringle}, J.~E. 1974, \mnras, 168, 603

\bibitem[{{Machida} {et~al.}(2000){Machida}, {Hayashi}, \& {Matsumoto}}]{mac00}
{Machida}, M., {Hayashi}, M.~R., \& {Matsumoto}, R. 2000, \apjl, 532, L67

\bibitem[{{Masada} {et~al.}(2012){Masada}, {Takiwaki}, {Kotake}, \&
  {Sano}}]{mas12}
{Masada}, Y., {Takiwaki}, T., {Kotake}, K., \& {Sano}, T. 2012, \apj, 759, 110

\bibitem[{{Matsumoto} \& {Tajima}(1995)}]{mt95}
{Matsumoto}, R., \& {Tajima}, T. 1995, \apj, 445, 767

\bibitem[{{McNally} \& {Pessah}(2015)}]{mp15}
{McNally}, C.~P., \& {Pessah}, M.~E. 2015, \apj, 811, 121

\bibitem[{{Miller} \& {Stone}(2000)}]{ms00}
{Miller}, K.~A., \& {Stone}, J.~M. 2000, \apj, 534, 398

\bibitem[{{Mohandas} \& {Pessah}(2017)}]{mp17}
{Mohandas}, G., \& {Pessah}, M.~E. 2017, \apj, 838, 48

\bibitem[{{Mori} {et~al.}(2019){Mori}, {Bai}, \& {Okuzumi}}]{mor19}
{Mori}, S., {Bai}, X.-N., \& {Okuzumi}, S. 2019, arXiv e-prints

\bibitem[{{Muto} {et~al.}(2010){Muto}, {Suzuki}, \& {Inutsuka}}]{mut10}
{Muto}, T., {Suzuki}, T.~K., \& {Inutsuka}, S.-i. 2010, \apj, 724, 448

\bibitem[{{Nakagawa} {et~al.}(1986){Nakagawa}, {Sekiya}, \& {Hayashi}}]{nak86}
{Nakagawa}, Y., {Sekiya}, M., \& {Hayashi}, C. 1986, ICARUS, 67, 375

\bibitem[{{Nelson} \& {Papaloizou}(2004)}]{np04}
{Nelson}, R.~P., \& {Papaloizou}, J.~C.~B. 2004, \mnras, 350, 849

\bibitem[{{Obergaulinger} {et~al.}(2009){Obergaulinger}, {Cerd{\'a}-Dur{\'a}n},
  {M{\"u}ller}, \& {Aloy}}]{obe09}
{Obergaulinger}, M., {Cerd{\'a}-Dur{\'a}n}, P., {M{\"u}ller}, E., \& {Aloy},
  M.~A. 2009, \aap, 498, 241

\bibitem[{{Ogihara} {et~al.}(2017){Ogihara}, {Kokubo}, {Suzuki}, {Morbidelli},
  \& {Crida}}]{ogi17}
{Ogihara}, M., {Kokubo}, E., {Suzuki}, T.~K., {Morbidelli}, A., \& {Crida}, A.
  2017, \aap, 608, A74

\bibitem[{{Okuzumi} \& {Hirose}(2012)}]{oh12}
{Okuzumi}, S., \& {Hirose}, S. 2012, \apjl, 753, L8

\bibitem[{{Okuzumi} {et~al.}(2014){Okuzumi}, {Takeuchi}, \& {Muto}}]{oku14}
{Okuzumi}, S., {Takeuchi}, T., \& {Muto}, T. 2014, \apj, 785, 127

\bibitem[{{Parkin} \& {Bicknell}(2013)}]{pb13}
{Parkin}, E.~R., \& {Bicknell}, G.~V. 2013, \apj, 763, 99

\bibitem[{{Pelletier} \& {Pudritz}(1992)}]{pp92}
{Pelletier}, G., \& {Pudritz}, R.~E. 1992, \apj, 394, 117

\bibitem[{{Penna} {et~al.}(2010){Penna}, {McKinney}, {Narayan}, {Tchekhovskoy},
  {Shafee}, \& {McClintock}}]{pen10}
{Penna}, R.~F., {McKinney}, J.~C., {Narayan}, R., {et~al.} 2010, \mnras, 408,
  752

\bibitem[{{Pringle}(1981)}]{pri81}
{Pringle}, J.~E. 1981, \araa, 19, 137

\bibitem[{{Rembiasz} {et~al.}(2016){Rembiasz}, {Guilet}, {Obergaulinger},
  {Cerd{\'a}-Dur{\'a}n}, {Aloy}, \& {M{\"u}ller}}]{rem16}
{Rembiasz}, T., {Guilet}, J., {Obergaulinger}, M., {et~al.} 2016, \mnras, 460,
  3316

\bibitem[{{Riols} {et~al.}(2016){Riols}, {Ogilvie}, {Latter}, \&
  {Ross}}]{rio16}
{Riols}, A., {Ogilvie}, G.~I., {Latter}, H., \& {Ross}, J.~P. 2016, \mnras,
  463, 3096

\bibitem[{{Rothstein} \& {Lovelace}(2008)}]{rl08}
{Rothstein}, D.~M., \& {Lovelace}, R.~V.~E. 2008, \apj, 677, 1221

\bibitem[{{Sano} {et~al.}(1999){Sano}, {Inutsuka}, \& {Miyama}}]{san99}
{Sano}, T., {Inutsuka}, S., \& {Miyama}, S.~M. 1999, in Astrophysics and Space
  Science Library, Vol. 240, Numerical Astrophysics, ed. S.~M. {Miyama},
  K.~{Tomisaka}, \& T.~{Hanawa} (Boston, MA: Kluwer), 383

\bibitem[{{Sano} {et~al.}(2004){Sano}, {Inutsuka}, {Turner}, \&
  {Stone}}]{san04}
{Sano}, T., {Inutsuka}, S.-i., {Turner}, N.~J., \& {Stone}, J.~M. 2004, \apj,
  605, 321

\bibitem[{{Sano} \& {Stone}(2002)}]{ss02}
{Sano}, T., \& {Stone}, J.~M. 2002, \apj, 570, 314

\bibitem[{{Shakura} \& {Sunyaev}(1973)}]{ss73}
{Shakura}, N.~I., \& {Sunyaev}, R.~A. 1973, \aap, 24, 337

\bibitem[{{Shi} {et~al.}(2010){Shi}, {Krolik}, \& {Hirose}}]{shi10}
{Shi}, J., {Krolik}, J.~H., \& {Hirose}, S. 2010, \apj, 708, 1716

\bibitem[{{Simon} {et~al.}(2018){Simon}, {Bai}, {Flaherty}, \&
  {Hughes}}]{sim18}
{Simon}, J.~B., {Bai}, X.-N., {Flaherty}, K.~M., \& {Hughes}, A.~M. 2018, \apj,
  865, 10

\bibitem[{{Simon} {et~al.}(2015){Simon}, {Lesur}, {Kunz}, \&
  {Armitage}}]{sim15}
{Simon}, J.~B., {Lesur}, G., {Kunz}, M.~W., \& {Armitage}, P.~J. 2015, \mnras,
  454, 1117

\bibitem[{{Stone} {et~al.}(1996){Stone}, {Hawley}, {Gammie}, \&
  {Balbus}}]{sto96}
{Stone}, J.~M., {Hawley}, J.~F., {Gammie}, C.~F., \& {Balbus}, S.~A. 1996,
  \apj, 463, 656

\bibitem[{{Suriano} {et~al.}(2019){Suriano}, {Li}, {Krasnopolsky}, {Suzuki}, \&
  {Shang}}]{sur19}
{Suriano}, S.~S., {Li}, Z.-Y., {Krasnopolsky}, R., {Suzuki}, T.~K., \& {Shang},
  H. 2019, \mnras, 484, 107

\bibitem[{{Suzuki} \& {Inutsuka}(2009)}]{si09}
{Suzuki}, T.~K., \& {Inutsuka}, S.-i. 2009, \apjl, 691, L49

\bibitem[{{Suzuki} \& {Inutsuka}(2014)}]{si14}
---. 2014, \apj, 784, 121

\bibitem[{{Suzuki} {et~al.}(2010){Suzuki}, {Muto}, \& {Inutsuka}}]{suz10}
{Suzuki}, T.~K., {Muto}, T., \& {Inutsuka}, S.-i. 2010, \apj, 718, 1289

\bibitem[{{Suzuki} {et~al.}(2016){Suzuki}, {Ogihara}, {Morbidelli}, {Crida}, \&
  {Guillot}}]{suz16}
{Suzuki}, T.~K., {Ogihara}, M., {Morbidelli}, A., {Crida}, A., \& {Guillot}, T.
  2016, \aap, 596, A74

\bibitem[{{Takahashi} \& {Muto}(2018)}]{tm18}
{Takahashi}, S.~Z., \& {Muto}, T. 2018, \apj, 865, 102

\bibitem[{{Takasao} {et~al.}(2018){Takasao}, {Tomida}, {Iwasaki}, \&
  {Suzuki}}]{tak18}
{Takasao}, S., {Tomida}, K., {Iwasaki}, K., \& {Suzuki}, T.~K. 2018, \apj, 857,
  4

\bibitem[{{Takeuchi} \& {Okuzumi}(2014)}]{to14}
{Takeuchi}, T., \& {Okuzumi}, S. 2014, \apj, 797, 132

\bibitem[{{Taki} {et~al.}(2016){Taki}, {Fujimoto}, \& {Ida}}]{tak16}
{Taki}, T., {Fujimoto}, M., \& {Ida}, S. 2016, \aap, 591, A86

\bibitem[{{Tanaka} {et~al.}(2002){Tanaka}, {Takeuchi}, \& {Ward}}]{tan02}
{Tanaka}, H., {Takeuchi}, T., \& {Ward}, W.~R. 2002, \apj, 565, 1257

\bibitem[{{Tanigawa} {et~al.}(2012){Tanigawa}, {Ohtsuki}, \& {Machida}}]{tan12}
{Tanigawa}, T., {Ohtsuki}, K., \& {Machida}, M.~N. 2012, \apj, 747, 47

\bibitem[{{Tomida} {et~al.}(2015){Tomida}, {Okuzumi}, \& {Machida}}]{tom15}
{Tomida}, K., {Okuzumi}, S., \& {Machida}, M.~N. 2015, \apj, 801, 117

\bibitem[{{Turner} {et~al.}(2007){Turner}, {Sano}, \& {Dziourkevitch}}]{tun07}
{Turner}, N.~J., {Sano}, T., \& {Dziourkevitch}, N. 2007, \apj, 659, 729

\bibitem[{{Turner} {et~al.}(2003){Turner}, {Stone}, {Krolik}, \&
  {Sano}}]{tur03}
{Turner}, N.~J., {Stone}, J.~M., {Krolik}, J.~H., \& {Sano}, T. 2003, \apj,
  593, 992

\bibitem[{{Velikhov}(1959)}]{vel59}
{Velikhov}, E.~P. 1959, Zh. Eksp. Teor. Fiz., 36, 1398

\bibitem[{{Yang} {et~al.}(2009){Yang}, {Mac Low}, \& {Menou}}]{yan09}
{Yang}, C.-C., {Mac Low}, M.-M., \& {Menou}, K. 2009, \apj, 707, 1233

\bibitem[{{Zhu} {et~al.}(2015){Zhu}, {Stone}, \& {Bai}}]{zhu15}
{Zhu}, Z., {Stone}, J.~M., \& {Bai}, X.-N. 2015, \apj, 801, 81

\end{thebibliography}

\end{document}